\documentclass[12pt]{elsarticle}
\usepackage{graphicx}
\usepackage{amssymb}    
\usepackage{epstopdf}
\usepackage{comment}
\usepackage{color}
\usepackage{hyperref}
\usepackage[final]{pdfpages}
\usepackage{frcursive}
\usepackage{bbm}
\usepackage{mathtools}
\usepackage{tabularx}
\usepackage{algorithm,algorithmic}

\usepackage{yfonts}
\usepackage{lscape}
\usepackage{epsfig}
\usepackage{fullpage}
\usepackage[titletoc]{appendix}
\usepackage{fancyhdr}
\usepackage{amsmath,amssymb,xspace}

\usepackage{tabularx}
\usepackage{mathrsfs}

\newfont{\bbb}{msbm10 scaled\magstep1}

\newtheorem{prop}{Proposition}[section]

\let \leq \leqslant
\let \geq \geqslant

{ \par \medskip \par
  \noindent \textit{\textbf{Demonstration\/}} : }{\null \hfill $\Box$ \par }
 
\newcommand{\R} {\ensuremath{\mathbb{R}}}
{

\makeatletter
\newcommand{\doublewidetilde}[1]{{%
  \mathpalette\double@widetilde{#1}%
}}
\newcommand{\double@widetilde}[2]{%
  \sbox\z@{$\m@th#1\widetilde{#2}$}%
  \ht\z@=.9\ht\z@
  \widetilde{\box\z@}%
}

\begin{document}

\begin{frontmatter}

\title{Inverse design of strained graphene surfaces for electron control}

\author[iqc]{Fran\c{c}ois Fillion-Gourdeau}
\ead{}

\author[carl,crm]{Emmanuel Lorin}
\ead{elorin@math.carleton.ca}

\author[iqc]{Steve MacLean}
\ead{}

\address[iqc]{Institute for Quantum Computing, University of Waterloo, Waterloo, Ontario, Canada, N2L~3G1.}
\address[carl]{School of Mathematics and Statistics, Carleton University, Ottawa, Canada, K1S 5B6}
\address[crm]{Centre de Recherches Math\'{e}matiques, Universit\'{e} de Montr\'{e}al, Montr\'{e}al, Canada, H3T~1J4}

\begin{abstract}
This paper is devoted to the inverse design of strained graphene surfaces for the control of electrons in the semi-classical optical-like regime. Assuming that charge carriers are described by the Dirac equation in curved-space and exploiting the fact that wave propagation can be described by ray-optics in this regime, a general computational strategy is proposed in order to find strain fields associated with a desired effective refractive index profile. The latter is first determined by solving semi-classical trajectories and by optimizing a chosen objective functional using a genetic algorithm. Then, the graded refractive index corresponding to the strain field is obtained by using its connection to the metric component in isothermal coordinates. These coordinates are evaluated via numerical quasiconformal transformations by solving the Beltrami equation with a finite volume method. The graphene surface deformation is finally optimized, also using a genetic algorithm, to reproduce the desired index of refraction. Some analytical results and numerical experiments are performed to illustrate the methodology.
\end{abstract}


\end{frontmatter}

\tableofcontents

\section{Introduction}

Straintronic, the control of electronic states by straining graphene and other 2D materials, has seen a surge of interest in the last decade because it promises new interesting physics \cite{Feng2016,GUINEA20121437,PhysRevLett.103.046801,Naumis_2017,AMORIM20161,PhysRevB.81.081407} and because it has  potential for applications, such as the Dirac fermion microscope \cite{boggild2017two}. When graphene is deformed or stretched, the interatomic distance is locally modified which in turn, changes the tight-binding description since the value of overlap integrals depends on the atomic position. Remarkably, this theoretical framework reduces to the 2D curved-space Dirac equation in the low-energy limit \cite{PhysRevB.76.165409,PhysRevLett.108.227205,PhysRevB.87.165131,OLIVALEYVA20152645,VOLOVIK2014352}, allowing for analogies between matter-gravity coupling theories and material science \cite{PhysRevB.82.073405,VOZMEDIANO2010109,Gallerati2019}.   

Understanding the behaviour of electrons in strained graphene is a challenging task, even in the low energy limit, because it requires a solution to the curved-space Dirac equation coupled to a pseudo-electromagnetic field. This equation has been solved in the time-independent case to characterize static properties of charge carriers \cite{PhysRevB.81.035408,Debus_2018,PhysRevB.95.125432,PhysRevB.84.081401}. In particular, this approach, along with other ones based on the tight-binding model \cite{RAMEZANIMASIR201376}, applied to homogeneously strained graphene has led to the discovery of Landau-like energy levels generated by large pseudo-magnetic fields that can reach up to 300 T \cite{Levy544}. The dynamic case, on the other hand, has not been investigated as thoroughly, in part because obtaining solutions to the time-dependent curved-space Dirac equation is more challenging (see \cite{CLY}). Nevertheless, some recent studies have tackled this challenge and demonstrated that wave packets can be manipulated by scattering on strained regions \cite{Chaves_2014,Debus_2018,Contreras_Astorga_2020}. For example, using numerical approaches, it was shown that electron wave packets can be confined \cite{PhysRevB.98.155419} or focused \cite{Stegmann_2016,PhysRevE.103.013312}.

In this work, we consider electron scattering over strained regions in the semi-classical and low-energy ($\lesssim 2$ eV) limit. The main goal is the inverse design of specific strain fields to steer and control charge carriers for applications in graphene nanoelectronics. To reach this goal, a set of numerical techniques is developed. A central part of our approach is the introduction of isothermal coordinates to describe the strained surface. The interest of working in isothermal coordinates is that its metric components can be interpreted physically as a graded index of refraction in the semi-classical approximation \cite{semiclassical2021}. The counterpart is that the construction of the metric tensor in isothermal coordinates requires the solution to the Beltrami equation, a first order system of partial differential equations. In this paper, this equation is numerically solved using a least-square cell-centered finite volume method, which offers a simple and flexible framework to solve partial differential equations with a reasonable accuracy. More importantly, however, is that it offers a direct connection between the strain field and the refractive index via semi-classical trajectories. We demonstrate that this feature can be exploited to inverse design strain fields by combining this approach with a standard metaheuristic optimization technique.   

The paper is organized as follows. In Section \ref{sec:strained_graph}, the curved-space Dirac equation in isothermal coordinates and its semiclassical limit are reviewed. In Section \ref{sec:summary}, we present the general strategy used for the inverse design of strained surfaces. Section \ref{sec:optics} is devoted to the optimization algorithm allowing to construct a desired graded index of refraction. A numerical scheme to solve the Beltrami equation is introduced in Section \ref{sec:beltrami}. We then propose  an original optimization method for parameterizing the  surface corresponding to the desired index of refraction in Section \ref{sec:iso}. In Section \ref{sec:numerics}, we propose some numerical experiments illustrating the overall strategy. We conclude in Section \ref{sec:conclusion}.

\section{Strained graphene in the semiclassical approximation} \label{sec:strained_graph}

In this section we review the mathematical model used to describe the dynamics of charge carriers in strained graphene. We also review its semiclassical approximation, which connects the geometry of the strained surface to an effective refractive index. 

\subsection{Curved-space Dirac equation}
Charge carriers in strained graphene are well-described by the massless Dirac equation in curved space-time at low energy \cite{OLIVALEYVA20152645,PhysRevLett.108.227205,Cortijo_2007,Vozmediano_2008,PhysRevB.88.085430,PhysRevB.87.165131,VOLOVIK2014352,Naumis_2017,AMORIM20161,VOZMEDIANO2010109,PhysRevB.88.155405}. This equation in covariant form reads \cite{PhysRevLett.108.227205,pollock2010dirac,PhysRevB.92.245110}
\begin{align}
\label{eq:dirac_cov}
 \texttt{i} \hbar \bar{\gamma}^{\mu}(q)D_{\mu}\psi(q) = 0,
\end{align} 
where $\psi(q)$ is the two-component spinor wave function, $q=(t,\boldsymbol{q})$ is a set of curvilinear coordinates (bold symbols are 2D vectors), $D_{\mu}$ is the curved-space covariant derivative and $\bar{\gamma}(q) = ( \bar{\gamma}^{0}(q),\bar{\gamma}^{i}(q) )$ are the generalized gamma matrices. The generalized gamma matrices are related to the metric of the surface via the local Clifford algebra
\begin{align}
\{\bar{\gamma}^{\mu}(q), \bar{\gamma}^{\nu}(q)\} = 2g^{\mu \nu} (q),
\end{align}
where $g^{\mu \nu}(q)$ is the metric of the space-time manifold.

One critical part of our approach is to write this equation for a general surface deformation using isothermal coordinates $\boldsymbol{r}$ where the metric is diagonal and yields a length element
\begin{eqnarray}
\label{eq:metric_iso}
ds^2 & = & v^2_Fdt^2 -\rho({\boldsymbol r})d{\boldsymbol r} \cdot d{\boldsymbol r} \, ,
\end{eqnarray}
where $v_{F} \approx c/300$ is Fermi's velocity in graphene and $\rho$ is the metric diagonal component.
In isothermal coordinates, the 2D massless curved-space static Dirac equation has a particularly simple form, reminiscent of the Dirac equation in flat space \cite{PhysRevE.103.013312}:
\begin{align}
\label{eq:dirac_iso}
{\tt i}\hbar \partial_t \psi(t,{\boldsymbol r}) &= 
-{\tt i}\frac{\hbar v_{F}}{\sqrt{\rho(\boldsymbol{r})}}\alpha^{i}   \left[\partial_{i} + \tilde{\Omega}_{i}({\boldsymbol r}) - {\tt i}A_{i}(\boldsymbol{r})\right]  
\psi(t,{\boldsymbol r}) \, ,
\end{align}
where $A_{i}(\boldsymbol{r})$ is the electromagnetic pseudo-vector potential and
where $\tilde{\Omega}_{i}(\boldsymbol{r}) = -\frac{1}{4} \partial_{i} \ln \big(\rho(\boldsymbol{r}) \bigr)$. The flat space Dirac matrices are given by $\alpha^{i} = \sigma^{i}$ (for $i=1,2$) and $\beta = \sigma^{3}$ ($\sigma^{i}$ are Pauli matrices). In graphene, the electromagnetic potential (responsible for pseudo-magnetic fields) and the spin connection appear naturally when the low energy limit of the tight-binding model is performed \cite{OLIVALEYVA20152645}.

Isothermal coordinates can be obtained for a given strained surface parametrized in Cartesian coordinates $\boldsymbol{x}$, by using quasi-conformal transformations characterized by the Beltrami equation \cite{ahlfors2006lectures,PhysRevE.103.013312}:
\begin{eqnarray}\label{beltrami}
P({\boldsymbol x})\nabla r_1({\boldsymbol x}) & = & J P({\boldsymbol x})\nabla r_2({\boldsymbol x}) \, ,
\end{eqnarray}
where 
\begin{eqnarray}\label{eqP}
P({\boldsymbol x}) = \cfrac{1}{\sqrt{1-|\mu({\boldsymbol x})|}}
\left[
\begin{array}{cc}
1-\mu_{\textrm{R}}({\boldsymbol x})& -\mu_{\textrm{I}}({\boldsymbol x}) \\
-\mu_{\textrm{I}}({\boldsymbol x}) & 1+\mu_{\textrm{R}}({\boldsymbol x})
\end{array}
\right], \, \, \, J & = & \left[
\begin{array}{cc}
0 & 1 \\ 
-1 & 0 
\end{array}
\right] \, ,
\end{eqnarray}
and 
\begin{eqnarray}\label{mu}
\mu({\boldsymbol x}) & = & \cfrac{E({\boldsymbol x})-G({\boldsymbol x})+2{\tt i}F({\boldsymbol x})}{E({\boldsymbol x}) +G({\boldsymbol x}) + 2\sqrt{E({\boldsymbol x})G({\boldsymbol x})-F^2({\boldsymbol x})}} \, .
\end{eqnarray}
This assumes that the 2D surface $\mathcal{S}$ embedded in a 3D space is parameterized by the displacement field 
\begin{align}
\label{eq:disp_field}
\vec{u}({\boldsymbol x})=(X({\boldsymbol x}),Y({\boldsymbol x}),Z({\boldsymbol x})),
\end{align} 
where the notation $\vec{v}$ stands for 3D vectors.
Then, the metric in Cartesian coordinates yields
\begin{align}
\label{eq:metric_cart}
ds^2  =  v^2_Fdt^2 -E({\boldsymbol r})dx^{2} - F({\boldsymbol r}) dxdy- G({\boldsymbol r})dy^{2},
\end{align}
with components 
\begin{eqnarray}\label{EFG}
\left\{
\begin{array}{lcl}
E(\boldsymbol{x}) = (\partial_{x}X)^{2} + (\partial_{x}Y)^{2} + (\partial_{x}Z)^{2}, \\
G(\boldsymbol{x}) = (\partial_{y}X)^{2} + (\partial_{y}Y)^{2} + (\partial_{y}Z)^{2}, \\
F(\boldsymbol{x}) = (\partial_{x}X)(\partial_{y}X) + (\partial_{x}Y)(\partial_{y}Y) + (\partial_{x}Z)(\partial_{y}Z)
\end{array}
\right. .
\end{eqnarray}
Once the Beltrami equation is solved, we can get the metric component in isothermal coordinates using
\begin{eqnarray}
\label{eq:rho}
	\rho({\boldsymbol x}) & = & \cfrac{E({\boldsymbol x}) + F({\boldsymbol x})+ 2\sqrt{E({\boldsymbol x})G({\boldsymbol x})-F^2({\boldsymbol x})}}{[\partial_x r_1({\boldsymbol x}) +\partial_y r_2({\boldsymbol x})]^2+[\partial_xr_2({\boldsymbol x})-\partial_y r_1({\boldsymbol x})]^2} \, .
\end{eqnarray}

\subsection{Semiclassical approximation}

As demonstrated in Ref. \cite{semiclassical2021}, the semi-classical limit of \eqref{eq:dirac_iso}, is evaluated by using the semi-classical ansatz
\begin{eqnarray}
\psi(t,\boldsymbol{r}) = e^{{\tt i}\frac{S(t,\boldsymbol{r})}{\hbar}} \sum_{n=0}^{\infty} \hbar^{n} u_{n}(t,\boldsymbol{r}),
\end{eqnarray}
where the amplitude $u$ is a bi-spinor and $S$ is a (real) phase \cite{maslov2001semi}. Collecting the $O(\hbar^{0})$ terms results in the following system of equations:
\begin{align}
\label{eq:real}
\left[\partial_{t}S(t,\boldsymbol{r}) +  \frac{v_{F} \alpha^{i}}{\sqrt{\rho(\boldsymbol{r})}}   \left[ \partial_{i} S(t,\boldsymbol{r}) \right]\right] u_{0}(t,\boldsymbol{r}) =0 ,
\end{align}
which can be solved by computing the determinant (in spinor-space). We get:
\begin{align}
\label{eq:eikonal}
\partial_{t}S(t,\boldsymbol{r}) =  h^{\pm}({\boldsymbol r},{\boldsymbol p}) = \pm \frac{v_{F}}{\sqrt{\rho(\boldsymbol{r})}} |\nabla S(t,\boldsymbol{r})|,
\end{align}
where $h^{\pm}$ is the classical Hamiltonian. Eq. \eqref{eq:eikonal} is the so-called eikonal equation for the curved-space Dirac equation. It can be expressed in Cartesian coordinates by performing a change of variable $\boldsymbol{r} \rightarrow \boldsymbol{x}$:
\begin{align}
\partial_{t}S(t,\boldsymbol{x}) = \pm \frac{v_{F}}{\sqrt{\rho(\boldsymbol{x})}} |\nabla S(t,\boldsymbol{x})|.
\end{align} 
Particle-like trajectories can be obtained from this equation via the method of characteristics. Letting $\boldsymbol{p} = \nabla S$, they are given explicitly by Hamilton's equations (we consider only positive solution $h = h^{+}$):
\begin{align}
\label{eq:eqm_x}
\frac{d \boldsymbol{x}}{dt} &= \nabla_{{\boldsymbol p}} h({\boldsymbol x},{\boldsymbol p}) =  \frac{v_{F}}{ n(\boldsymbol{x})}  \frac{\boldsymbol{p}(t,\boldsymbol{x})}{|\boldsymbol{p}(t,\boldsymbol{x})|}, \\
\label{eq:eqm_p}
\frac{d \boldsymbol{p}}{dt} &= -\nabla_{{\boldsymbol x}} h({\boldsymbol x},{\boldsymbol p}) =  v_{F} \left( \frac{\nabla n(\boldsymbol{x})}{n^{2}(\boldsymbol{x})}\right) | \boldsymbol{p}(t,\boldsymbol{x})| ,
\end{align}
where we defined $n(\boldsymbol{x}) = \sqrt{\rho(\boldsymbol{x})}$. This function $n(\boldsymbol{x})$ can be interpreted as a graded index of refraction by evaluating the speed using Eq. \eqref{eq:eqm_x}: $|\boldsymbol{v}(\boldsymbol{x})|=|\dot{\boldsymbol{x}}| = \frac{v_{F}}{n(\boldsymbol{x})}$. Thus, the metric component in isothermal coordinates has a direct effect on wave propagation.

Finally, it is possible to express the equation of motion in a Newton-like form, by taking the time-derivative of the velocity \eqref{eq:eqm_x}. We find
\begin{align}
\ddot{\boldsymbol{x}} = \frac{v_{F}^{2}}{ n^{3}(\boldsymbol{x})}\nabla n(\boldsymbol{x}).
\end{align}
Changing the time variable from $t$ to $a$, such that $|d\boldsymbol{x}/da| = n(\boldsymbol{x})$ (following Evans' formulation \cite{Newton}), the equation of motion becomes
\begin{align}
\label{eq:newton}
\begin{cases}
\cfrac{d^{2} \boldsymbol{x}}{d a} = \nabla \left[\cfrac{n^{2}(\boldsymbol{x})}{2} \right] \\
\Big|\cfrac{d{\boldsymbol x}(0)}{da}\Big|  =  n({\boldsymbol x}_0) \\
{\boldsymbol x}(0)  =  {\boldsymbol x}_0 
\end{cases} .
\end{align}
Thus, the semi-classical trajectories can be evaluated from this differential equation. Physically, these trajectories are important because they are orthogonal to wavefronts.

\section{Control of charge carriers in graphene}\label{sec:summary}
The main objective of this paper is to use computational methods to design strained graphene surfaces in the semi-classical regime which passively control the trajectories of electrons. Such control is interesting from an application point of view because it allows for quantum lensing or waveguide effects \cite{BARTELMANN2001291,PhysRevB.82.205430,PhysRevB.78.235321,doi:10.1021/nn800459e,PhysRevB.81.081407,Chaves_2014,RevModPhys.81.109}. The typical configuration under consideration is a free incoming electron wave packet that scatters on a locally deformed region.  

In order to achieve this goal, we assume that the system is in the semi-classical regime, where typical deformations vary slowly compared to the electron wave function. Then, we can benefit from the isomorphism between the index of refraction $n(\boldsymbol{x})$ and the surface parametrization given in Eq. \eqref{eq:disp_field} via the solution of the Beltrami equation \eqref{beltrami} and the metric component in isothermal coordinates \eqref{eq:rho}. Exploiting this connection, we proceed in two steps:
\begin{enumerate}
\item A desired graded refractive index $n(\boldsymbol{x})$ is chosen.
\item The displacement field $\vec{u}(\boldsymbol{x})$ corresponding to the desired graded index profile is determined.
\end{enumerate}

The first step is to choose a particular index profile such that electronic rays, that obey the classical-like equation \eqref{eq:newton}, are directed in some given direction or follow some specific trajectories. This is a classic inverse problem in transformation optics \cite{LEONHARDT200969} and is solved here using a metaheuristic algorithm, as described in Section \ref{sec:optics}.  

The second step is to determine the displacement field that corresponds to the desired index profile. It is challenging to obtain the displacement field $\vec{u}(\boldsymbol{x})$ and deformations $X,Y$ and $Z$ associated to a specific strain-induced refractive index profile. Mathematically, this requires the inversion of Eq. \eqref{eq:rho}, which  non-linearly depends on the solution of the Beltrami equation \eqref{beltrami}. Although this inversion is very challenging and may even be impossible analytically in the general case, it can accurately be performed numerically by solving  an inverse minimization problem. A numerical scheme to solve \eqref{beltrami} is proposed in Section \ref{sec:beltrami} while a metaheuristic algorithm is used in Section \ref{sec:iso} to find the displacement field.


\section{Optimization of the graded refractive index profile}\label{sec:optics}

In this section, we discuss the construction of the desired graded refractive index. The specific index profile depends on the considered application. Here, we are specifically interested in designing graphene surfaces that behave like gravitational lenses \cite{ga1,ga2} where each ``electron-ray'' (usually referred to as ray hereafter) is focused to a single point. 

More specifically, we consider $\mathcal{R}>1$ rays, $\{\boldsymbol x_i\}_{1\leq i\leq \mathcal{R}}$ satisfying Eq. \eqref{eq:newton} with distinct initial conditions $\{\boldsymbol x_{0;i}\}_{1\leq i\leq \mathcal{R}}$. Our objective is to determine a parametrized graded index of refraction $n_{\pi}$, where the $v$-dimensional parameter vector $\pi$ belongs to a bounded search space $\Pi \subset \R^{v}$, such that the rays intersect at (or close to) a given target point ${\boldsymbol x}_T$. We also want this point to be reached at the same ``final time'' $\sigma$ (which will also be an optimization parameter). The objective function minimization hence reads
\begin{eqnarray}\label{opt_opt}
\textrm{argmin}_{\pi;\sigma}\sum_{i=1}^\mathcal{R}\|{\boldsymbol x}_i(\sigma)-{\boldsymbol x}_T\|_2 \, .
\end{eqnarray}
In other words, we need to simultaneously optimize the function $\rho$, but also the ``crossing stepping time'' denoted by $\sigma$.

The rays are evaluated numerically. We denote by ${\boldsymbol X}^s_i$ a finite difference approximation of ${\boldsymbol x}_i(a_k)$, where $a_0=0,a_1,\cdots,a_k,\cdots$ with $a_k=k\Delta a$. For instance, we can consider the {\it second order} approximation, for $k>1$
\begin{eqnarray}\label{ODEsolver}
{\boldsymbol X}^{k+1}_i-2{\boldsymbol X}^{k}_i + {\boldsymbol X}^{k-1}_i & = & \cfrac{\Delta a^2}{2}\nabla n^{2}_{\pi}({\boldsymbol X}^{k}_i) \, ,
\end{eqnarray} 
with ${\boldsymbol X}^0_i={\boldsymbol x}_{0;i}$. For $k=1$ and using the initial condition, a natural approximation is 
\begin{eqnarray*}
	{\boldsymbol X}^{1}_i = {\boldsymbol x}_{0;i} +(\Delta a)  n({\boldsymbol x}_{0;i}) + \cfrac{\Delta a^2}{2}\nabla n^{2}_{\pi}({\boldsymbol x}_{0;i}) \, .
\end{eqnarray*} 
The minimization of the discrete cost function hence reads (with $N_{\sigma}\Delta a=\sigma$),
\begin{eqnarray}
	\label{eq:opt_dis}
	\mathcal{F} = \textrm{argmin}_{\pi \in \Pi; 1\leq k\leq N_{\sigma}}\sum_{i=1}^\mathcal{R}\|{\boldsymbol X}^k_{i}-{\boldsymbol x}_T\|_2 \, .
\end{eqnarray}
The problem presented in Eq. \eqref{eq:opt_dis} is a single-objective optimization problem on continuous variables in a $(v+1)$-dimension hypercube search space. Metaheuristic algorithms are particularly well-suited for these types of problem \cite{yang2010nature}. Accordingly, an evolutionary algorithm (EA) is chosen to explore the parameter space and find a solution close to a minimum. 

In a nutshell, EAs are population-based and proceed as follows. At the beginning, a number of random ``individuals'' are created forming the first generation. Each individual corresponds to one point in parameter space with a specific set of parameters $(\pi,\sigma)$. Then, a sequence of new population is generated, where the fittest individuals are more likely to be passed to the next generation. The fitness value is directly related to the value of the objective function: in our case, higher fitness is associated to a lower value of $\mathcal{F}$. These individuals are then randomly modified (mutation) or combined (crossover) to create new individuals in the population. New generations are created until some stopping criterion is reached.  Individuals with the highest fitness are then selected as champions.  

There exists several variants of EAs. In this article, the standard population-based genetic algorithm \cite{ga1,ga2} implemented in \textsc{Matlab} is chosen, in which individual mutations are performed by adding a random Gaussian distributed vector while the crossovers are accomplished by a random weighted average of the parents.

\subsection{Numerical example: Gaussian index profile}
\begin{figure}[hbt!]
	\begin{center}
		\includegraphics[height=6cm,keepaspectratio]{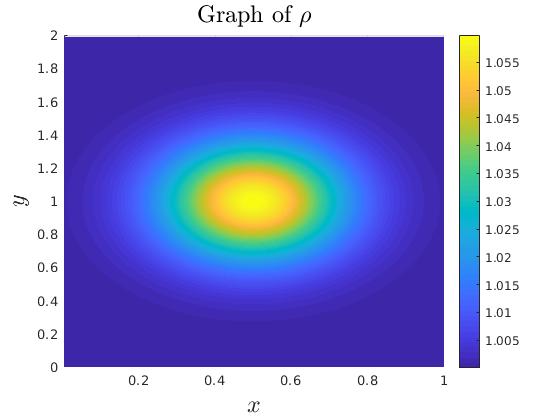}
		\includegraphics[height=6cm,keepaspectratio]{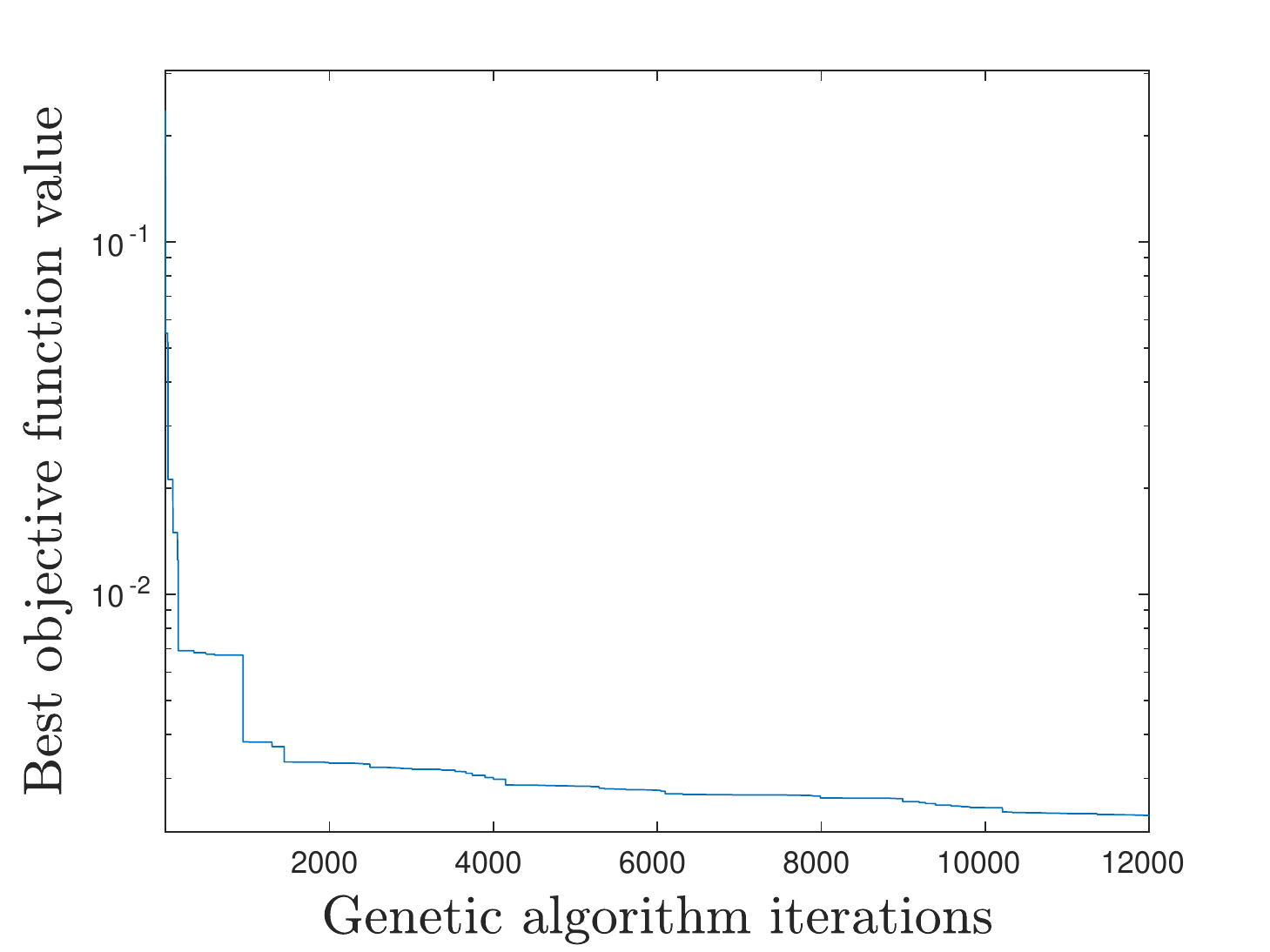}
	\end{center}
	\caption{(Left) Graph of optimized $\rho = n^{2}$. (Right) Best discrete objective function as a function of total number of iterations of the genetic algorithm.}
	\label{fig2}
\end{figure}

\begin{figure}[hbt!]
\begin{center}
\includegraphics[height=5.5cm,keepaspectratio]{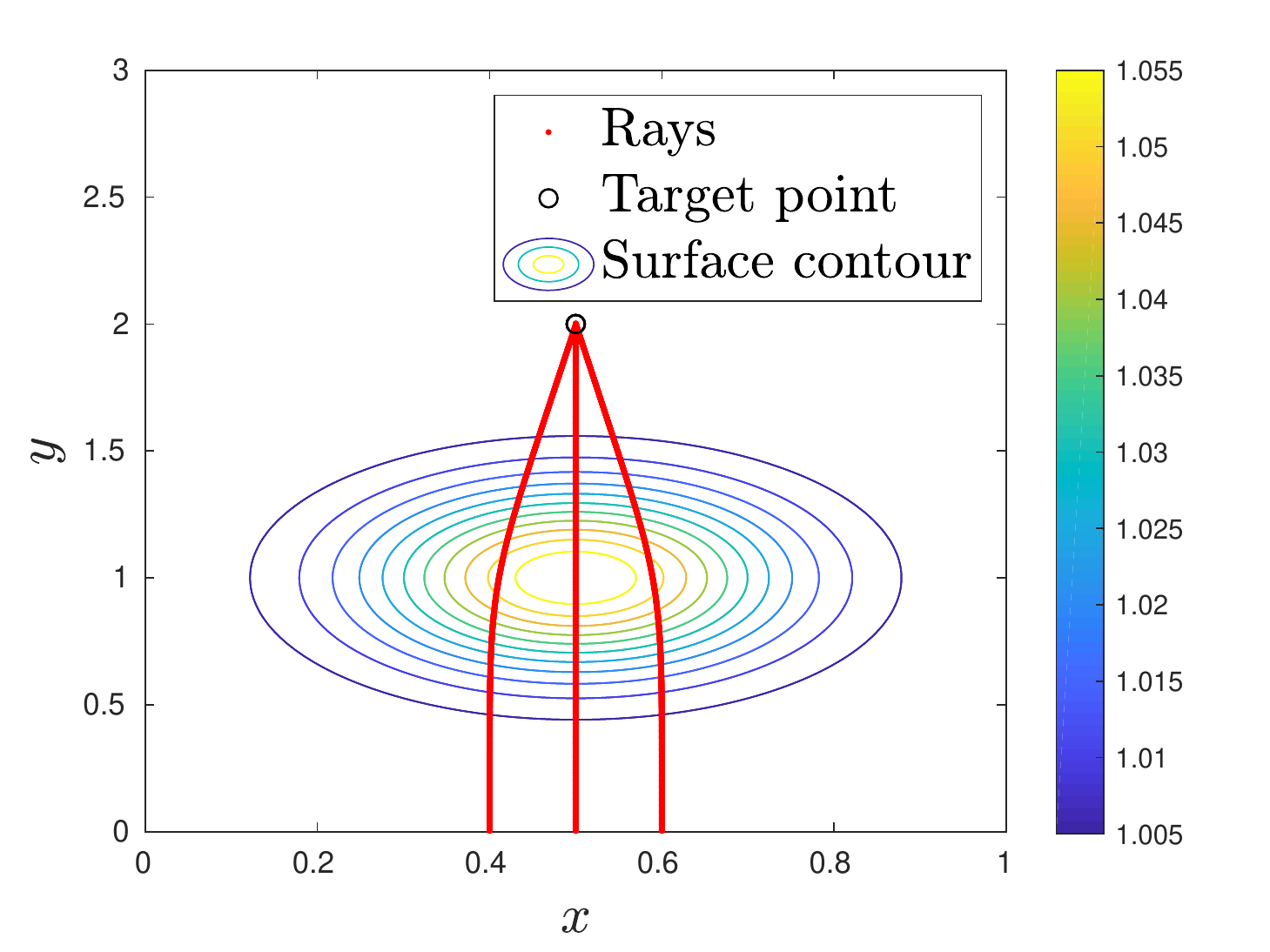}
\includegraphics[height=5.5cm,keepaspectratio]{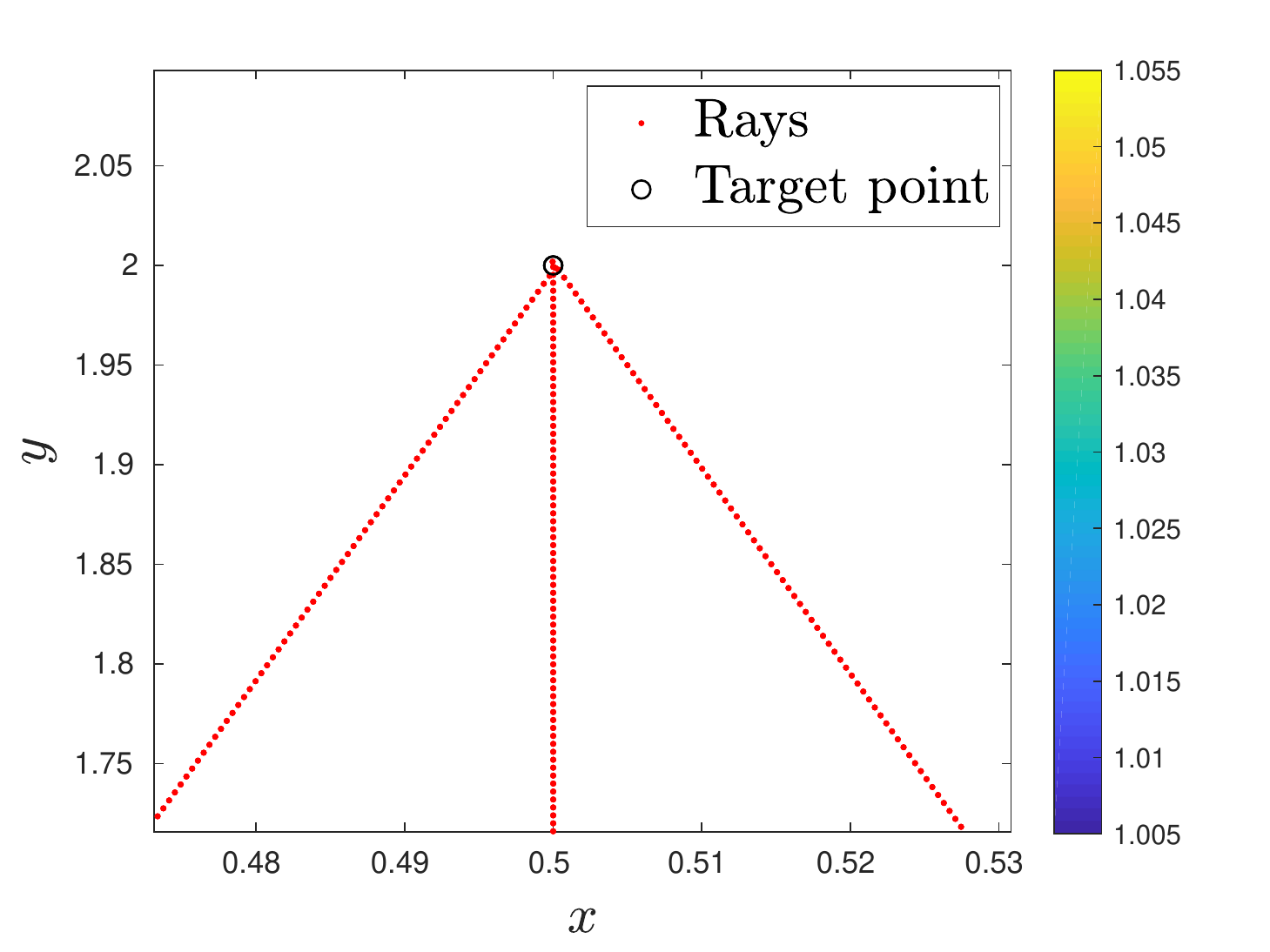}
\end{center}
\caption{(Left) Semi-classical trajectories of the champion (Right) Close-up on the trajectories in the vicinity of the target point.}
\label{fig1}
\end{figure}

In this first numerical experiment, we consider 3 rays, initially located at ${\boldsymbol x}_{0;i}=(0.4+0.1i,0)$ ($i=1,2,3$). We assume that the index profile is a Gaussian function parameterized by $\pi=(A,w_x,w_y)$ such that
\begin{eqnarray*}
n^{2}_{\pi}({\boldsymbol x}) & = & 1+ A\exp\big(-w_x(x-0.5)^2-w_y(y-1)^2\big) .
\end{eqnarray*}
The target point is selected as ${\boldsymbol x}_T=(0.5,2)$ and the $4$-dimensional search space is given by $\big(\pi,\sigma\big) \in [0.05,0.15]\times[2,20]\times[2,10]\times[1.9,2.1]$. When one of the stopping criteria is fulfilled, the genetic algorithm provides the following champion  $(\pi^{*},\sigma^{*})=(A,w_x,w_y,\sigma)=(0.0599,17.3786,7.9505,1.9948)$. In Fig. \ref{fig2}, the graph of $\rho_{\pi^*} = n^{2}_{\pi^*}$ is displayed along with the logarithm of the fittest individual as a function of the total number of iterations of the genetic algorithm. We also report the semi-classical trajectories of the champion in Fig. \ref{fig1}. This figure demonstrates that the trajectories are intersecting at the target point.

\section{Solution to the Beltrami equation}\label{sec:beltrami}

As argued in Section \ref{sec:summary}, the Beltrami equation \eqref{beltrami} needs to be solved to evaluate the displacement field. A least-square finite-element method was proposed in Ref. \cite{PhysRevE.103.013312} to reach this goal. In this section, we derive a simple second-order accurate least-square cell-centered finite volume method. 

\subsection{Least-square finite volume framework}\label{subsec:LSFVS}

We introduce a conforming finite volume mesh $\Omega_h=\cup_{j=1}^{N_{j}}K_j$ covering the domain $\Omega$, where the volumes $K_j$ are typically chosen as rectangles or triangles \cite{MR1925043}. The polygonal boundary of $\Omega_h$ is denoted by $\Gamma_h$. We also use the following notation:
\begin{itemize} 
\item the edges of a finite volume $K_j$ are denoted $\{e_{j;i}\}_{i=1}^{e}$, where $e$ ($=3,4$) is the number of edges.
\item  the outward norm vector to the edge $e_{j;i}$ is denoted by ${\boldsymbol n}_{ji}$ and  $d\sigma_{j;i}({\boldsymbol x})$ (practically Lebesgue's measure) denotes the measure along $e_{j;i}$, 
\item the area of the volume $K_i$ is denoted $|K_i|$, and the length of $e_{j;i}$ is denoted by $|e_{j;i}|$,
\item the volume having the edge $e_{j;i}$ in common with $K_j$ is denoted $K_{j_i}$, where $1\leq j_i\leq N_{j}$ and $j_i\neq j$.
\end{itemize}
In order to solve \eqref{beltrami}, we search for cell-center finite volume functions $R_1$, $R_2$  of the form
\begin{eqnarray*}
R_1({\boldsymbol x}) =  \sum_{j=1}^{N_{j}}a_j{\bf 1}_{K_j}({\boldsymbol x}), \, \, \, \, \,  R_2({\boldsymbol x}) =  \sum_{j=1}^{N_{j}}b_j{\bf 1}_{K_j}({\boldsymbol x}) \, ,
\end{eqnarray*}
where $\{a_j\}_{j}$ and  $\{b_j\}_{j}$ are the approximate values of $r_1$ and $r_2$ on the volume $\{K_j\}_j$, and ${\bf 1}_K$ denotes the characteristic function on $K$. We also denote by $a_{j;i}$ (resp. $b_{j;i}$) the approximate values of $r_1$ (resp. $r_2$) on $e_{j;i}$. Then, we integrate \eqref{beltrami} over $\Omega_h$. The left-hand-side reads
\begin{align}
\label{eq:bel_fv_lhs}
\int_{\Omega_h}P({\boldsymbol x})\nabla r_1({\boldsymbol x})d{\boldsymbol x} & = \sum_{j=1}^{N_{j}}\int_{K_j}P({\boldsymbol x})\nabla r_1({\boldsymbol x})d{\boldsymbol x} \nonumber \\
& =  \sum_{j=1}^{N_{j}}\sum_{i=1}^{e}\int_{e_{j;i}}P({\boldsymbol x}) {\boldsymbol n}_{j;i}r_1({\boldsymbol x})d \sigma_{j;i}({\boldsymbol x})-\sum_{j=1}^{N_{j}}\int_{K_j}\nabla P({\boldsymbol x}) \otimes r_1({\boldsymbol x})d{\boldsymbol x}\, ,
\end{align}
while the right-hand-side reads
\begin{align}
\label{eq:bel_fv_rhs}
	\int_{\Omega_h}Q({\boldsymbol x})\nabla r_2({\boldsymbol x})d{\boldsymbol x} & = \sum_{j=1}^{N_{j}}\int_{K_j}Q({\boldsymbol x})\nabla r_2({\boldsymbol x})d{\boldsymbol x} \nonumber  \\
	& =  \sum_{j=1}^{N_{j}}\sum_{i=1}^{e}\int_{e_{j;i}}Q({\boldsymbol x}) {\boldsymbol n}_{j;i}r_2({\boldsymbol x})d \sigma_{j;i}({\boldsymbol x})-\sum_{j=1}^{N_{j}}\int_{K_j}\nabla Q({\boldsymbol x}) \otimes r_2({\boldsymbol x})d{\boldsymbol x}\, ,
\end{align}
where $Q = JP$ and where we have denoted 
\begin{eqnarray}\label{FVSTOT}
\nabla P({\boldsymbol x}) \otimes r_1({\boldsymbol x}) & = & 
\left[
\begin{array}{l}
\partial_x P_{11} + \partial_y P_{12}  \\
\partial_x P_{21} + \partial_y P_{22}
\end{array}
\right] r_1({\boldsymbol x})\, .
\end{eqnarray}
We next denote for any matrix valued function ${\boldsymbol T}$
\begin{eqnarray*}
T_{j;i} = \cfrac{1}{|e_{j;i}|}\int_{e_{j;i}}{\boldsymbol T}({\boldsymbol x})d\sigma_{j;i}({\boldsymbol x}), \, \, \, \, \nabla T_{j} =  \cfrac{1}{|K_j|}\int_{K_j}\nabla {\boldsymbol T}({\boldsymbol x})d{\boldsymbol x} \, .
\end{eqnarray*}

\subsection{Finite volume approximation for interior volumes}\label{subsec:FVS}

In this subsection, we focus on the finite volume approximation for volumes not having a common edge with the boundary $\Gamma_h$ (designated as interior volumes). 
From \eqref{eq:bel_fv_lhs} and \eqref{eq:bel_fv_rhs}, we then propose the following finite volume approximation on each interior volume $K_j$ :
\begin{eqnarray*}
\sum_{i=1}^e|e_{j;i}|P_{j;i}{\boldsymbol n}_{j;i} a_{j;i} - |K_j|\nabla P_{j}\otimes a_j & = & \sum_{i=1}^e|e_{j;i}|Q_{j;i}{\boldsymbol n}_{j;i} b_{j;i} - |K_j|\nabla Q_{j}\otimes b_j \, .
\end{eqnarray*}
In order to get an explicit expression of the scheme as a function of $\{a_j\}_{1\leq j \leq J}$ and  $\{b_j\}_{1\leq j\leq N_{j}}$, we also propose a standard approximation of the edge values, as follows
\begin{eqnarray*}
a_{j;i}  =  \cfrac{a_j |K_j| + a_{j_i}|K_{j_i}| }{|K_{j_i}|+|K_j|}, \, \, \, \, \,  b_{j;i} = \cfrac{b_j |K_j| + b_{j_i}|K_{j_i}| }{|K_{j_i}|+|K_j|} \, ,
\end{eqnarray*}
where $|K_{j_i}|$ is the area of  the neighboring volume $K_{j_i}$ to $K_j$, having $e_{j;i}$ as a common edge. The finite volume scheme, hence reads for each interior volume: search for $\{a_j\}_j$, and $\{b_j\}_j$ such that
\begin{eqnarray}\label{FVScheme}
\left.
\begin{array}{lcl}
\sum_{i=1}^e|e_{j;i}|P_{j;i}{\boldsymbol n}_{j;i}  \cfrac{a_j |K_j| + a_{j_i}|K_{j_i}| }{|K_{j_i}|+|K_j|} - |K_j|\nabla P_{j}\otimes a_j & = & \sum_{i=1}^eQ_{j;i}{\boldsymbol n}_{j;i}  \cfrac{b_j |K_j| + b_{j_i}|K_{j_i}| }{|K_{j_i}|+|K_j|} \\
& &  - |K_j|\nabla Q_{j}\otimes b_j \, .
\end{array}
\right.
\end{eqnarray}
The above expression is hence an algebraic system with $\{a_j\}_{1\leq j \leq N_{j}}$ and $\{b_j\}_{1\leq j \leq N_{j}}$ as unknowns. 
\subsection{Boundary conditions}
We now detail the treatment of Dirichlet boundary conditions. For $k=1,2$, we impose
\begin{eqnarray*}
r_k({\boldsymbol x}) & = & g_k({\boldsymbol x}), \, \, \textrm{ on } \Gamma \, ,
\end{eqnarray*}
where $g_k$ are given functions. For finite volumes $K_j$ sharing an edge $e_{j;i}$ with $\Gamma_h$ (that is $e_{j;i}\subset \Gamma_h$), we approximate  $a_{j;i}$ by
\begin{eqnarray*}
a_{j;i} & \simeq & \cfrac{a_j + g_{1;j}}{2} \, ,
\end{eqnarray*}
where $g_{1;j}$ is the mean of (an extension of) $g_1$ on a {\it ghost} volume (cell), symmetric to $K_j$ with respect to $e_{j;i}$. 
\\
\\
In fine, the finite volume scheme can simply be written in the form
\begin{eqnarray}\label{FVS}
\mathcal{K}_a{\boldsymbol a} +  {\boldsymbol F}_a   = \mathcal{K}_b{\boldsymbol b} + {\boldsymbol F}_b \, ,
\end{eqnarray}
where i) ${\boldsymbol a} \in \R^{N_{j}}$ and ${\boldsymbol b}  \in \R^{N_{j}}$ are the unknown coefficients of $R_1$ and $R_2$, ii) $\mathcal{K}_{a,b} \in \R^{N_{j}\times N_{j}}$, and iii)  ${\boldsymbol F}_{a,b}\in \R^{N_{j}}$ are the boundary condition contributions in \eqref{FVScheme}.

In order to solve \eqref{FVS}, we use a standard least-square method. More specifically, we need to compute 
\begin{eqnarray*}
\label{FVS_min}
\textrm{argmin}_{({\boldsymbol a}, {\boldsymbol b}) \in\R^{N_{j}}\times\R^{N_{j}}}\|\mathcal{K}_a{\boldsymbol a}-\mathcal{K}_b{\boldsymbol b} + {\boldsymbol F}_a - {\boldsymbol F}_b\|^2 \, .
\end{eqnarray*}
The least-square problem is finally solved using a quasi-newton function minimizer, using the {\tt matlab} function called {\tt fminunc}.

\subsection{Mathematical analysis of the least-square finite-volume method}
In this subsection, we present some analytical properties of the least-square finite-volume method introduced in this paper. More specifically, we first focus on the order of consistency. Let us consider a smooth function $f$ defined on a two-dimensional rectangle finite volume $K=[-\Delta x/2,\Delta x/2]\times[-\Delta y/2,\Delta y/2]$ of area $|K|=\Delta x\Delta y$ and centered at ${\boldsymbol 0}$. We denote by $f_{K}$ the mean function of $f$ over $K$. We prove the following result on the accuracy of our finite-volume method for a flat surface (with $P=1$) and for a curved surface. Note that in the former, the Beltrami equation degenerates into the Cauchy-Riemann equation.
\begin{prop}
The approximation of the Cauchy-Riemann (resp. Beltrami on a smooth surface) equation with Dirichlet boundary conditions, using the cell-center finite volume method \eqref{FVS_min} with rectangle cells is second (resp. first) order accurate.
\end{prop}
{\bf Proof.} First, we recall that for $K=[-\Delta x/2,\Delta x/2]\times [-\Delta y/2,\Delta y/2]$
\begin{eqnarray*}
\left.
\begin{array}{lcl}
f_{K}  =  \cfrac{1}{|K|}\int_{K}f({\boldsymbol x})d{\boldsymbol x}= \cfrac{1}{\Delta x\Delta y}\int_{-\Delta x/2}^{\Delta x/2}\int_{-\Delta y/2}^{\Delta y/2}f(x,y)dxdy = f({\boldsymbol 0}) + O(\Delta x^2+\Delta y^2) \, .
\end{array}
\right.
\end{eqnarray*}
Hence by Taylor's expansion of $f$ about ${\boldsymbol 0}$, we also get for ${\boldsymbol x} \in K$
\begin{eqnarray*}
f({\boldsymbol x}) & = & f_{K} + O(\Delta x+\Delta y) \ . 
\end{eqnarray*}
 Let us denote by $L_{i}$ ($i=1,2,3,4$) the neighboring volume sharing the edge $e_i=K\cap L_i$ with $K$, such that ${\boldsymbol n}_{KL_{1}}=-{\boldsymbol n}_{KL_{3}}=(1,0)^T$, and ${\boldsymbol n}_{KL_{2}}=-{\boldsymbol n}_{KL_{4}}=(0,1)^T$. By a slight change of notation, we will also denote by $f_{L_i}$ the constant value of the function $f$ in the volume $L_i$. Then
\begin{eqnarray*}
\left.
\begin{array}{lcl}
\int_{\cup_{i=1}^4e_i} f({\boldsymbol x}){\boldsymbol n} d{\sigma}({\boldsymbol x}) & = & \int_{-\Delta x/2}^{\Delta x/2} f\big(x,\frac{\Delta y}{2}\big)-f\big(x,-\frac{\Delta y}{2}\big)dx  + \int_{-\Delta y/2}^{\Delta y/2} f\big(\frac{\Delta x}{2},y\big)-f\big(-\frac{\Delta x}{2},y\big)dy \\
& = &  \int_{-\Delta x/2}^{\Delta x/2} f\big(x,\frac{\Delta y}{2}\big)-f\big(x,-\Delta y/2)dx +  \int_{-\Delta y/2}^{\Delta y/2} f\big(\frac{\Delta x}{2},y\big)-f\big(-\frac{\Delta x}{2},y\big)dy \\
& = &  \int_{-\Delta x/2}^{\Delta x/2} \partial_y f(x,0) dx +\int_{-\Delta y/2}^{\Delta y/2} \partial_x f(0,y) dy +  O(\Delta x \Delta y^2+ \Delta y \Delta x^2) \\
& = & \Delta x \partial_y f({\boldsymbol 0}) +  \Delta y \partial_x f({\boldsymbol 0}) +  O(\Delta x \Delta y^2+ \Delta y \Delta x^2 +\Delta x^3+ \Delta y^3) \\
& = & \frac{\Delta x}{2\Delta y}(f_{L_3}+f_{L_1}) + \frac{\Delta y}{2\Delta x}(f_{L_4}+f_{L_2}) +  O(\Delta x\Delta y^2 + \Delta y \Delta x^2 +\Delta x^3+ \Delta y^3) \, .
\end{array}
\right.
\end{eqnarray*}
The latter equation is hence second-order consistent with \eqref{FVS_min}.

In the case of the Beltrami equation, we theoretically loose one order of accuracy due to the contribution of the smooth matrices $P$ and $Q$ (we skip the details which are trivially a consequence of standard Taylor's expansions). The first order term is hence only proportional to $\|\nabla P\|_{\infty}$ (as $\|\nabla Q\|_{\infty}=\|\nabla P\|_{\infty}$), and more specifically to $\|\nabla P\|_{\infty}(\Delta x + \Delta y)$. This concludes the proof. $\Box$\\

It is interesting to notice that for Gaussian surfaces (Beltrami equation case), as $P$ is constituted by Gaussian-like functions, the first order term coefficient is proportional to the sup-norm of the derivative of Gaussian functions, so that an order $2$-like behavior is still expected.
\subsection{Numerical examples}

We propose in this subsection, a series of numerical experiments to illustrate the proposed least-square finite volume method.

\subsubsection{Flat surface} 

\begin{figure}[hbt!]
\begin{center}
\includegraphics[height=5.5cm,keepaspectratio]{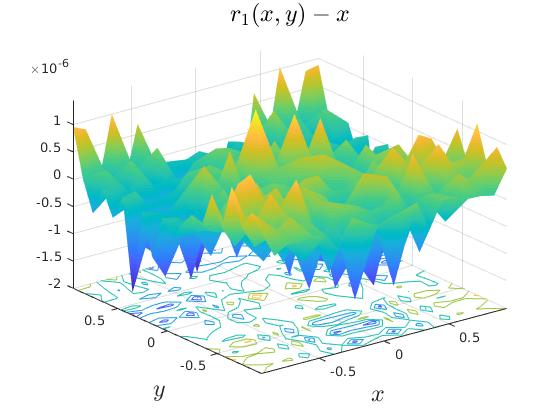}
\includegraphics[height=5.5cm,keepaspectratio]{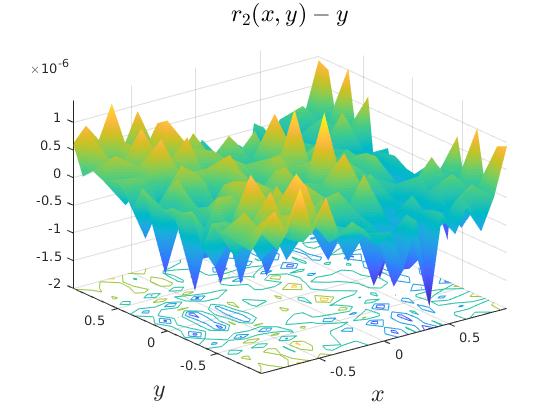}
\end{center}
\caption{{\bf Experiment 2.} (Left) Graph of error function for $r_1$. (Right).  Graph of error function for $r_1$.}
\label{figCR}
\end{figure}

In the following experiment, we consider the case of a flat surface  $\Omega = [x_{\mathrm{min}},x_{\mathrm{max}}] \times [y_{\mathrm{min}},y_{\mathrm{max}}]$. As mentioned above, the Beltrami equation degenerates into the Cauchy-Riemann equation
\begin{eqnarray}\label{CR}
\nabla r_1({\boldsymbol x}) & = & J \nabla r_2({\boldsymbol x}) \, ,
\end{eqnarray}
with the following boundary conditions
\begin{align}
\label{eq:bel_bound_u}
\left. r_1 \right|_{\partial \Omega^{\mathrm{l}}} &= x_{\mathrm{min}}, & 
\left. r_1 \right|_{\partial \Omega^{\mathrm{r}}} &= x_{\mathrm{max}} ,  \nonumber \\ 
\left. r_1 \right|_{\partial \Omega^{\mathrm{d}}} &= y, &
\left. r_1 \right|_{\partial \Omega^{\mathrm{t}}} &= y ,& \\
\label{eq:bel_bound_v}
\left. r_2 \right|_{\partial \Omega^{\mathrm{l}}} &= x, &
\left. r_2 \right|_{\partial \Omega^{\mathrm{r}}} &= x , \nonumber \\ 
\left. r_2 \right|_{\partial \Omega^{\mathrm{d}}} &= y_{\mathrm{min}},& 
\left. r_2 \right|_{\partial \Omega^{\mathrm{t}}} &= y_{\mathrm{max}} ,&
\end{align}
where $\partial \Omega^{\mathrm{l,r,d,t}}$ are the left, right, down and top boundaries of the rectangular domain, respectively. In this case, the exact solution is given by $r_1({\boldsymbol x})=x$ and $r_2({\boldsymbol x})=y$. \\
The test is performed on $\Omega = [-1,1]^2$ with an initial guess given by $r_1({\boldsymbol x})=r_2({\boldsymbol x})=1$, and we report the error functions ($r_1({\boldsymbol x})-x$, $r_2({\boldsymbol x})-y$) obtained with $20^2$ volumes on Fig. \ref{figCR}. The corresponding error $\ell^2$-error $\|R_1-x\|_2 + \|R_2-y\|_2$ is $2 \times 10^{-11}$ and the $\ell^{\infty}$-norm is given by $3\times 10^{-6}$.

\subsubsection{Curved Gaussian surface} 

\begin{figure}[hbt!]
\begin{center}
\includegraphics[height=5.5cm,keepaspectratio]{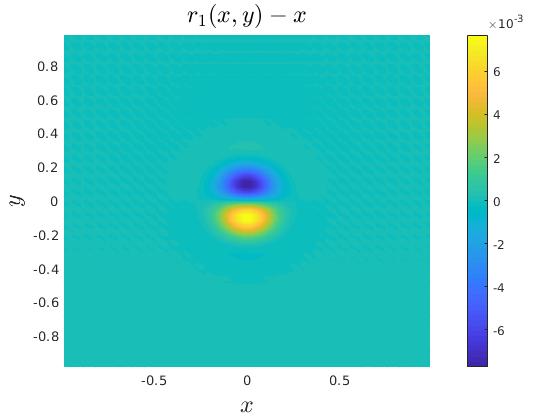}
\includegraphics[height=5.5cm,keepaspectratio]{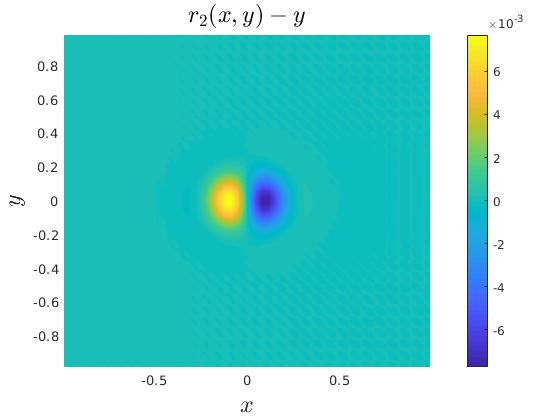}
\end{center}
\caption{(Left) Graph of $r_1({\boldsymbol x})-y$ (Right).  Graph of $r_2({\boldsymbol x})-y$.}
\label{figB}
\end{figure}

In this experiment, we assume that the surface is defined from the displacement field $\vec{u}({\boldsymbol x})=({\boldsymbol x},Z({\boldsymbol x}))$ with $Z({\boldsymbol x})=10^{-1}\exp\big(-10\|{\boldsymbol x}-{\boldsymbol c}\|^2\big)$, and $\Omega=[-1,1]^2$. We report the graph of $r_1({\boldsymbol x})-x$ and $r_2({\boldsymbol x})-y$, with $60^2$ finite volumes on Fig. \ref{figB}. The solution is consistent with \cite{PhysRevE.103.013312}.

\section{Evaluation of the displacement field}\label{sec:iso}

In this section, we develop a method to find the displacement field associated to a target refractive index profile. The starting point is the surface $\mathcal{S}_{\Lambda}$ characterized by the following displacement field $\vec{u}_{\Lambda}({\boldsymbol x})=(X_{\Lambda}({\boldsymbol x}),Y_{\Lambda}({\boldsymbol x}),Z_{\Lambda}({\boldsymbol x}))$. Here, $\Lambda$ denotes a set of parameters that parametrizes the strained surface. These parameters are optimized in a search space denoted by $\mathcal{L}$. Then, we set the target index of refraction to $\sqrt{\rho_{T}(\boldsymbol{x})} = n_{T}(\boldsymbol{x}) = n_{\pi}(\boldsymbol{x})$, where $n_{\pi}$ is the refractive index obtained in Section \ref{sec:optics}. Therefore, to obtained the corresponding strained surface, we minimize the cost function
\begin{eqnarray*}
\textrm{argmin}_{\Lambda \in \mathcal{L}}\|\rho_T-\rho_{\Lambda}\|_2 , \\
\begin{matrix}
	\mbox{Subject to:}\;&\mathcal{C}_{I} \leq 0 \\
	&\mathcal{C}_{E} = 0 \, ,
\end{matrix}
\end{eqnarray*}
where the computation of $\rho_{\Lambda}$ for given a $\Lambda$, is performed via the solution to a Beltrami equation as explained in Section \ref{sec:beltrami} or in \cite{PhysRevE.103.013312}.  Notice in particular, that the construction of $\rho_{\Lambda}$ we will require to compute intermediate functions $E_{\Lambda},F_{\Lambda},G_{\Lambda}$ as in \eqref{EFG}.  

The vectors $\mathcal{C}_{I}$ and $\mathcal{C}_{E}$ are sets of constraints. Their explicit form will depend on the displacement field parametrization and on the physical configuration which is considered. For example, one physical constraint that should always be taken into account is that the strain should never be larger than the maximum strain that graphene can sustain $ \epsilon_{\mathrm{max}} \approx 0.25$ \cite{Lee385}. This can be formulated as
\begin{align}
\mathcal{C}_{I,0} = \max_{\boldsymbol{x}} \left[ \mathcal{E}(\boldsymbol{x})\right] - \epsilon_{\mathrm{max}} \leq 0,
\end{align} 
where $\mathcal{E}$ is the norm of some strain measure. 
Obviously, other constraints can be implemented to faithfully represent physical or experimental limitations.

In order to keep a total flexibility on the size of the searched space containing $\Lambda$, we propose to use the same genetic algorithm as in Section \ref{sec:optics}. The overall scheme is summarized in Algorithm 1.

\begin{algorithm}
\caption{Optimization problem for the displacement field}
\begin{algorithmic}[1]
\STATE Define the bounded search space $\mathcal{L}$ and the constraints.
\STATE Compute $\rho_{\Lambda}$ numerically for $\Lambda \in \mathcal{L}$.
\begin{itemize}
\item For $\Lambda \in \mathcal{L}$, compute the solution to the Beltrami equation (see details in Section \ref{sec:beltrami})
\item Function $\rho_{\Lambda}$ is given by equation \eqref{eq:rho}
\end{itemize}
\STATE Estimate the cost function $\|\rho_T-\rho_{\Lambda}\|_2$.
\STATE Until a stopping criterion is reached, stochastically update $\Lambda \in \mathcal{L}$.
\end{algorithmic}
\end{algorithm}

\subsection{Mathematical analysis}


We are interested in the design of a graphene surface having a desired index of refraction to control charge carriers. However, there is no guarantee that the optimization algorithm will find the global minimum, rather it will only capture local minima. Therefore, in the first proposition, we prove that if the surfaces $\mathcal{S}_T$ and $\mathcal{S}_{\Lambda}$ are close enough, their metric in isothermal coordinates will also be close. Here, the index $T$ (resp. $\Lambda$) is used to refer to the target (resp. estimated or optimized) quantities $\rho_T$,  $\mathcal{S}_T=\{({\boldsymbol x},Z_T({\boldsymbol x}))\}$, $E_T$, etc (resp. $\rho_{\Lambda}$,  $\mathcal{S}_{\Lambda}=\{({\boldsymbol x},Z_{\Lambda}({\boldsymbol x}))\}$, $E_{\Lambda}$, etc). This proposition is proven to ensure that even if the global minimum is not found, we obtain a displacement field that is close to the correct one.

We denote by $r_i^{T}$ (resp. $r^{\Lambda}_i$) the solution to \eqref{BelTilde} on $\mathcal{S}_T$ (resp. $\mathcal{S}_{\Lambda}$).
\begin{prop}
	Assume that there exists a small $\varepsilon>0$, $M_T>0$, $M_{\Lambda}>0$,  such that $\|\nabla Z-\nabla Z_{\Lambda}\|_{\infty}<\varepsilon$, $\|\nabla Z_{T}\|_{\infty}<M_T$, $\|\nabla Z_{\Lambda}\|_{\infty}<M_{\Lambda}$. Moreover, for $i=1,2$, we denote by $R_{T,L}$ (resp. $r_{T,L}$) an upper (lower) bound of $\nabla r_i^{T,L}$, that is
	\begin{eqnarray*}
		r_T \leq \|\nabla r_i^T\|_{\infty} \leq R_T, \, \, \, \, \, \, r_{\Lambda} \leq \|\nabla r_i^{\Lambda}\|_{\infty} \leq R_{\Lambda} \, .
	\end{eqnarray*}
	Then there exists a constant $C=C(M_T,M_{\Lambda},R_T,r_T)>0$, such that
	\begin{eqnarray*}
		\|\rho_T-\rho_{\Lambda}\|_{\infty} & < & C\varepsilon  \, .
	\end{eqnarray*}
\end{prop}
{\bf Proof.} We consider the Beltrami equation written in the form 
\begin{eqnarray}\label{BelTilde}
\widetilde{P}({\boldsymbol x})\nabla r_1({\boldsymbol x}) & = & J \widetilde{P}({\boldsymbol x})\nabla r_2({\boldsymbol x}) \, ,
\end{eqnarray}
where 
\begin{eqnarray}\label{Ptilde}
\widetilde{P}({\boldsymbol x}) = \left[
\begin{array}{cc}
1-\mu_{\textrm{R}}({\boldsymbol x})& -\mu_{\textrm{I}}({\boldsymbol x}) \\
-\mu_{\textrm{I}}({\boldsymbol x}) & 1+\mu_{\textrm{R}}({\boldsymbol x})
\end{array}
\right], \, \, \, J & = & \left[
\begin{array}{cc}
0 & 1 \\ 
-1 & 0 
\end{array}
\right] \, .
\end{eqnarray}
The proof requires several intermediate estimates. We first notice that
\begin{eqnarray*}
	\left.
	\begin{array}{lcl}
		\|E_T-E_{\Lambda}\|_{\infty} & = & \|(\partial_x Z_T)^2 - (\partial_x Z_{\Lambda})^2\|_{\infty} \\
		& = & \|\partial_x Z_T + \partial_x Z_{\Lambda}\|\|\partial_x Z_T - \partial_x Z_{\Lambda}\|_{\infty} \\
		& \leq & (M_T+M_{\Lambda})\varepsilon \, .
	\end{array}
	\right.
\end{eqnarray*}
The same results hold for $\|G_T-G_{\Lambda}\|_{\infty}$.  Regarding $\|F_T-F_{\Lambda}\|_{\infty}$, we notice that
\begin{eqnarray*}
	\left.
	\begin{array}{lcl}
		\|F_T-F_{\Lambda}\|_{\infty} & = & \|(\partial_x Z_T)(\partial_y Z_T) - (\partial_x Z_{\Lambda})^2(\partial_y Z_{\Lambda})\|_{\infty} \\
		& = & \|(\partial_x Z_T)(\partial_y Z_T) - (\partial_x Z_T)(\partial_y Z_{\Lambda}) + (\partial_x Z_T)(\partial_y Z_{\Lambda})- (\partial_x Z_{\Lambda})(\partial_y Z_{\Lambda})\|_{\infty}\\
		& \leq & (M_T+M_{\Lambda})\varepsilon \, .
	\end{array}
	\right.
\end{eqnarray*}
Next, we  remark that for any ${\boldsymbol x}$, as $E_{T}({\boldsymbol x})\geq 1$, $E_{\Lambda}({\boldsymbol x})\geq 1$, $G_T({\boldsymbol x})\geq 1$, $G_{\Lambda}({\boldsymbol x})\geq 1$, we easily deduce that there exists $C(M_T,M_{\Lambda})>0$ such that
\begin{eqnarray}\label{estimate_mu}
\|\mu_T-\mu_{\Lambda}\|_{\infty} & \leq & C\varepsilon \, .
\end{eqnarray}
We now focus on error estimates related to the modified Beltrami equation \eqref{BelTilde}. Let us denote $s_i=R^T_i-r^{\Lambda}_i$ for $i=1,2$. Then, from \eqref{BelTilde} for $r_i^T$ and $r^{\Lambda}_i$ we get
\begin{eqnarray*}
	\left.
	\begin{array}{lcl}
		\nabla s_1({\boldsymbol x}) & = & \widetilde{P}_{\Lambda}^{-1}({\boldsymbol x})J \widetilde{P}({\boldsymbol x})\nabla s_2({\boldsymbol x})\\
		& &  + \widetilde{P}_{\Lambda}^{-1}({\boldsymbol x})J \big(\widetilde{P}_T({\boldsymbol x}) - \widetilde{P}_{\Lambda}({\boldsymbol x})\big)\nabla r^T_2({\boldsymbol x}) - \widetilde{P}_{\Lambda}^{-1}({\boldsymbol x})\big(\widetilde{P}_T({\boldsymbol x}) - \widetilde{P}_{\Lambda}({\boldsymbol x})\big) \nabla r^T_1({\boldsymbol x})\, .
	\end{array}
	\right.
\end{eqnarray*}
As $E_{\Lambda}({\boldsymbol x}) \geq 1$, $G_{\Lambda}({\boldsymbol x})\geq 1$ and using \eqref{estimate_mu}, we easily show that there exists a constant $C=C(M_T,M_{\Lambda},R_T)>0$ such that $\|\widetilde{P}\|_{\infty}<C\varepsilon$, so that
\begin{eqnarray*}
	\|\widetilde{P}_{\Lambda}^{-1}J \big(\widetilde{P}_T - \widetilde{P}_{\Lambda}\big)\nabla r^T_2 - \widetilde{P}_{\Lambda}^{-1}\big(\widetilde{P}_T - \widetilde{P}_{\Lambda}\big) \nabla r^T_1\|_{\infty} \leq C\varepsilon \, .
\end{eqnarray*}
Moreover, as the solution to 
\begin{eqnarray*}
	\widetilde{P}({\boldsymbol x})\nabla s_1({\boldsymbol x}) & = & J \widetilde{P}({\boldsymbol x})\nabla s_2({\boldsymbol x}) \, ,
\end{eqnarray*}
with null Dirichlet boundary conditions is null, we easily deduce that there exists $C=C(M_T,M_{\Lambda},R_T)>0$ such that
\begin{eqnarray}\label{ErrBel}
\|\nabla r_i^T - \nabla r^{\Lambda}_i \| & \leq & C \varepsilon \, .
\end{eqnarray}
Finally, setting $\Delta \rho = \rho_{T}-\rho_{\Lambda}$, we get
\begin{eqnarray*}
	\left.
	\begin{array}{lll}
		\Delta \rho = \cfrac{\big([\partial_x r^{\Lambda}_1 +\partial_y r^{\Lambda}_2]^2+[\partial_xr^{\Lambda}_2-\partial_y r^{\Lambda}_1]^2\big)\big(E_{T} + F_{T}+ 2\sqrt{E_{T}G_{T}-F_{T}^2}\big)}{\big([\partial_x r^{T}_1 +\partial_y r^{T}_2]^2+[\partial_xr^{T}_2-\partial_y r^{T}_1]^2\big)\big([\partial_x r^{\Lambda}_1 +\partial_y r^{\Lambda}_2]^2+[\partial_xr^{\Lambda}_2-\partial_y r^{\Lambda}_1]^2\big)} \\
		-\cfrac{ \big([\partial_x r^{T}_1 +\partial_y r^{T}_2]^2+[\partial_xr^{T}_2-\partial_y r^{T}_1]^2\big)\big(E_{\Lambda} + F_{\Lambda}+ 2\sqrt{E_{\Lambda}G_{\Lambda}-F_{\Lambda}^2}\big)}{\big([\partial_x r^{T}_1 +\partial_y r^{T}_2]^2+[\partial_xr^{T}_2-\partial_y r^{T}_1]^2\big)\big([\partial_x r^{\Lambda}_1 +\partial_y r^{\Lambda}_2]^2+[\partial_xr^{\Lambda}_2-\partial_y r^{\Lambda}_1]^2\big)} \\
		=  \cfrac{\big([\partial_x r^{\Lambda}_1 +\partial_y r^{\Lambda}_2]^2+[\partial_xr^{\Lambda}_2-\partial_y r^{\Lambda}_1]^2\big)\big(E_{T} + F_{T}+ 2\sqrt{E_{T}G_{T}-F_{T}^2} - E_{\Lambda} - F_{\Lambda}- 2\sqrt{E_{\Lambda}G_{\Lambda}-F_{\Lambda}^2}\big)}{\big([\partial_x r^{T}_1 +\partial_y r^{T}_2]^2+[\partial_xr^{T}_2-\partial_y r^{T}_1]^2\big)\big([\partial_x r^{\Lambda}_1 +\partial_y r^{\Lambda}_2]^2+[\partial_xr^{\Lambda}_2-\partial_y r^{\Lambda}_1]^2\big)}\\
		+ \cfrac{ \big([\partial_x r^{\Lambda}_1 +\partial_y r^{\Lambda}_2]^2+[\partial_xr^{\Lambda}_2-\partial_y r^{\Lambda}_1]^2-[\partial_x r^{T}_1 +\partial_y r^{T}_2]^2-[\partial_xr^{T}_2-\partial_y r^{T}_1]^2\big)\big(E_{\Lambda} + F_{\Lambda}+ 2\sqrt{E_{\Lambda}G_{\Lambda}-F_{\Lambda}^2}\big)}{\big([\partial_x r^{T}_1 +\partial_y r^{T}_2]^2+[\partial_xr^{T}_2-\partial_y r^{T}_1]^2\big)\big([\partial_x r^{\Lambda}_1 +\partial_y r^{\Lambda}_2]^2+[\partial_xr^{\Lambda}_2-\partial_y r^{\Lambda}_1]^2\big)} \, ,
	\end{array}
	\right.
\end{eqnarray*}
and combining estimates of $\|E_T-E_{\Lambda}\|_{\infty}$, $\|G_T-G_{\Lambda}\|_{\infty}$ and \eqref{ErrBel}, we deduce the existence of a constant $C=C(M_t,M_{\Lambda},R_t,r_T)>0$ such that
\begin{eqnarray*}
	\|\rho_T-\rho_{\Lambda}\|_{\infty} & < & C\varepsilon  \, .
\end{eqnarray*}
This concludes the proof. $\Box$
\subsection{Numerical example: Gaussian target function}
\begin{figure}
\begin{center}
\includegraphics[height=4cm,keepaspectratio]{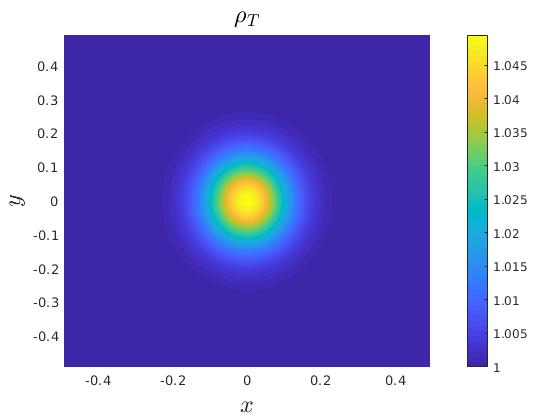}
\includegraphics[height=4cm,keepaspectratio]{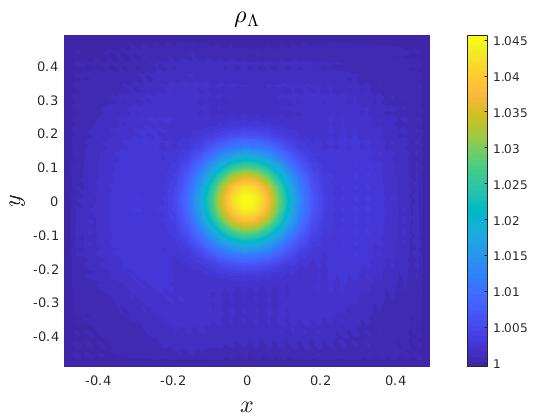}
\includegraphics[height=4cm,keepaspectratio]{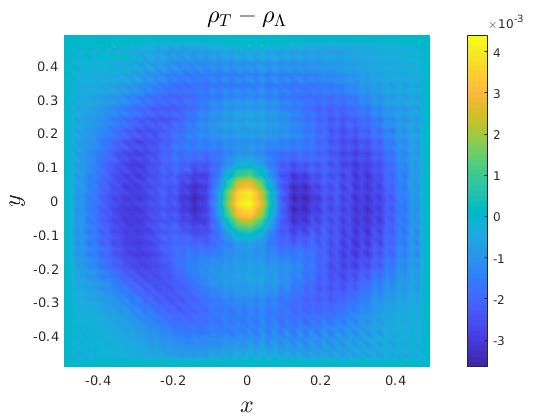}
\end{center}
\caption{(Left) Graph of the target function $\rho_T$. (Middle) Graph of the optimized function $\rho_{\Lambda^*}$ (Right) Error function.}
\label{opt1}
\end{figure}

\begin{figure}[hbt!]
\begin{center}
\includegraphics[height=6cm,keepaspectratio]{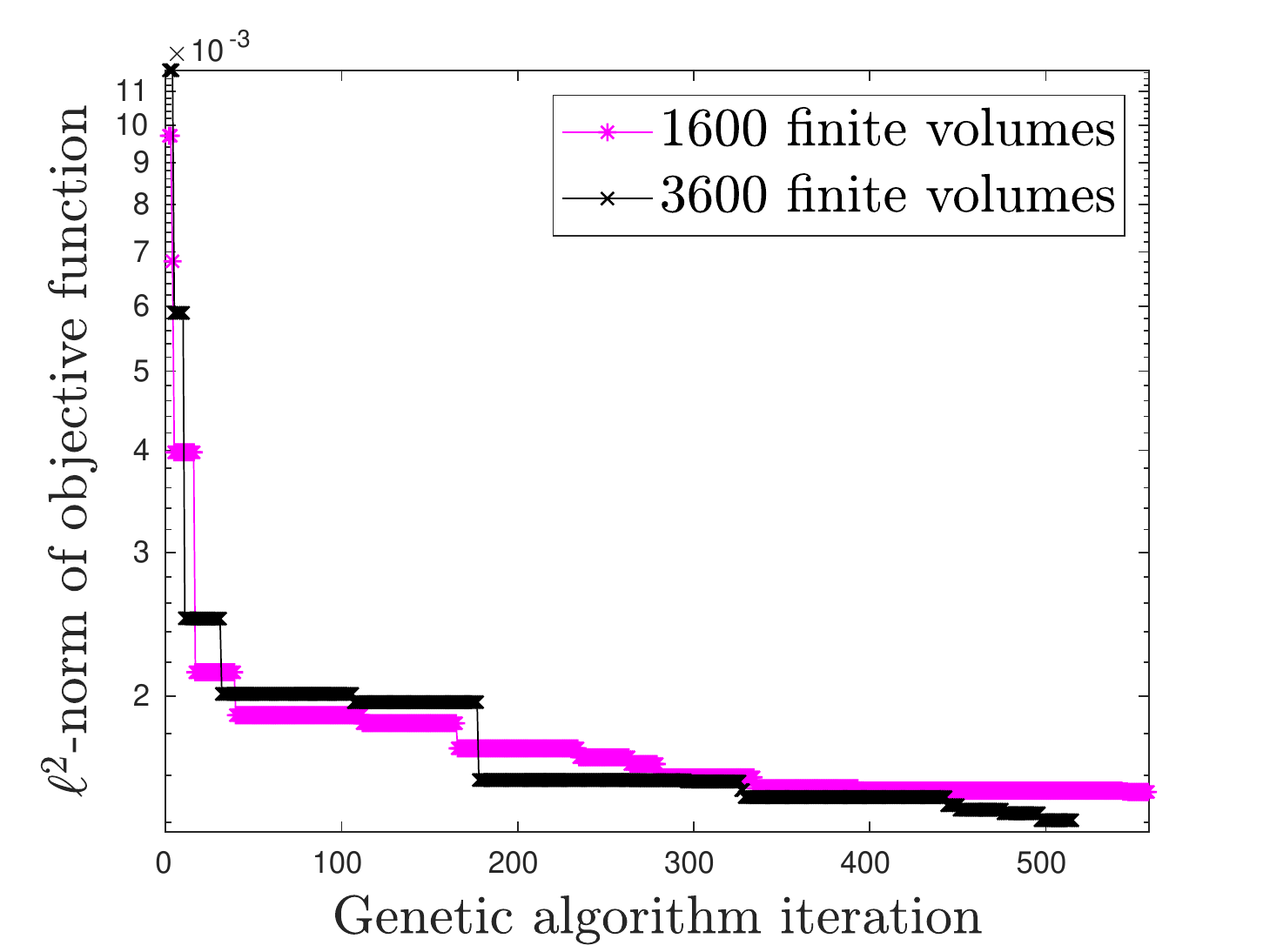}
\includegraphics[height=6cm,keepaspectratio]{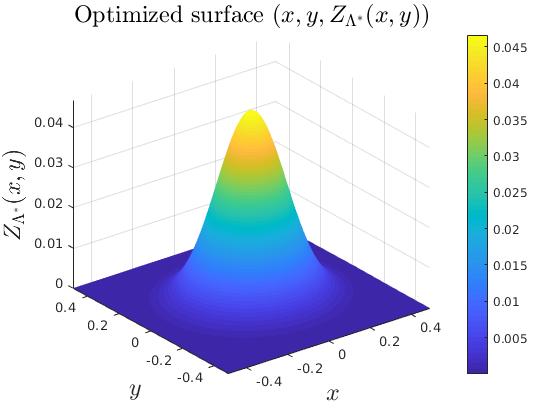}
\end{center}
\caption{(Left) Best discrete objective function as a function of total number of iterations of the genetic algorithm for $1600$ and $3600$ finite volumes.  (Right) Optimized surface.}
\label{opt2}
\end{figure}

We propose a simple test illustrating the optimization scheme summarized in Algorithm 1. We assume that the domain is $[-1/2,1/2]^2$ and we fix the target function to
\begin{eqnarray}
\rho_T(\boldsymbol{x}) = c_1\exp(-c_2x^2-c_3y^2),
\end{eqnarray}
where $c_1=1/20$, $c_2=60$, $c_3=80$. To reproduce this target function, the displacement field is parametrized as $u_{\Lambda}({\boldsymbol x})=({\boldsymbol x},Z_{\Lambda}({\boldsymbol x}))$, where 
\begin{eqnarray*}
Z_{\Lambda}({\boldsymbol x}) & = & A\exp\big(-w_xx^2-w_y y^2\big) \, .
\end{eqnarray*}
Thus, the 3 unknown parameters $\Lambda=(A,w_x,w_y)$ are the optimization parameters. We use a total of $1600$ (then $3600$) square finite volumes on which i) we project the target function $\rho_{T}$, and ii) we solve the Beltrami equation. Initially, we take $\Lambda=(0.01,5,5)$ and the 3-dimensional search space is $\mathcal{L}=[0.005,1.1]\times[1,45]\times [1,45]$.  We report in Fig. \ref{opt1} (Left) the target function, (Middle) the optimized functions, and (Right) the error function $\rho_T-\rho_{\Lambda}$. The local minima obtained with the code are $\Lambda^*=(A^*,w^*_x,w^*_y)=(0.0409,19.2634,21.8627)$, for 1600 finite volumes (and $(0.0468,12.6860,14.9177)$ for 3600 finite volumes). We report in Fig. \ref{opt2} (Left) the objective functions as a function of the total number of iterations of the optimization algorithm.
The corresponding surface $({\boldsymbol x},Z_{\Lambda^{*}}({\boldsymbol x}))$ with ${\boldsymbol x} \in [-1/2,1/2]^2$ is finally reported in Fig. \ref{opt2} (Right). We see that the method and the surface parametrization are able to reproduce the target function with a relatively small error $O(10^{-3})$. We also see that a finer discretization reduces the error.

\section{Experiments in refractive optics on strained graphene surfaces}\label{sec:numerics}

In this section, we propose two complete tests in which a desired index of refraction is determined from the technique of Section \ref{sec:optics} and the corresponding surface is obtained from the method in Section \ref{sec:iso}.

\subsection{Lens without aberration}

The main goal of this example is to produce an aberration-free lens, i.e. a lens where all the rays meet at the same target point. To design a strained surface with such an effect on charge carriers, five rays are considered and are initially located at ${\boldsymbol x}_{0;i}=(0.4+0.05i,0)$. Physically, they represent the wavefront of an incoming wave packet propagating at a certain velocity in the $y$-direction. 

\begin{figure}[hbt!]
\begin{center}
\includegraphics[height=5.5cm,keepaspectratio]{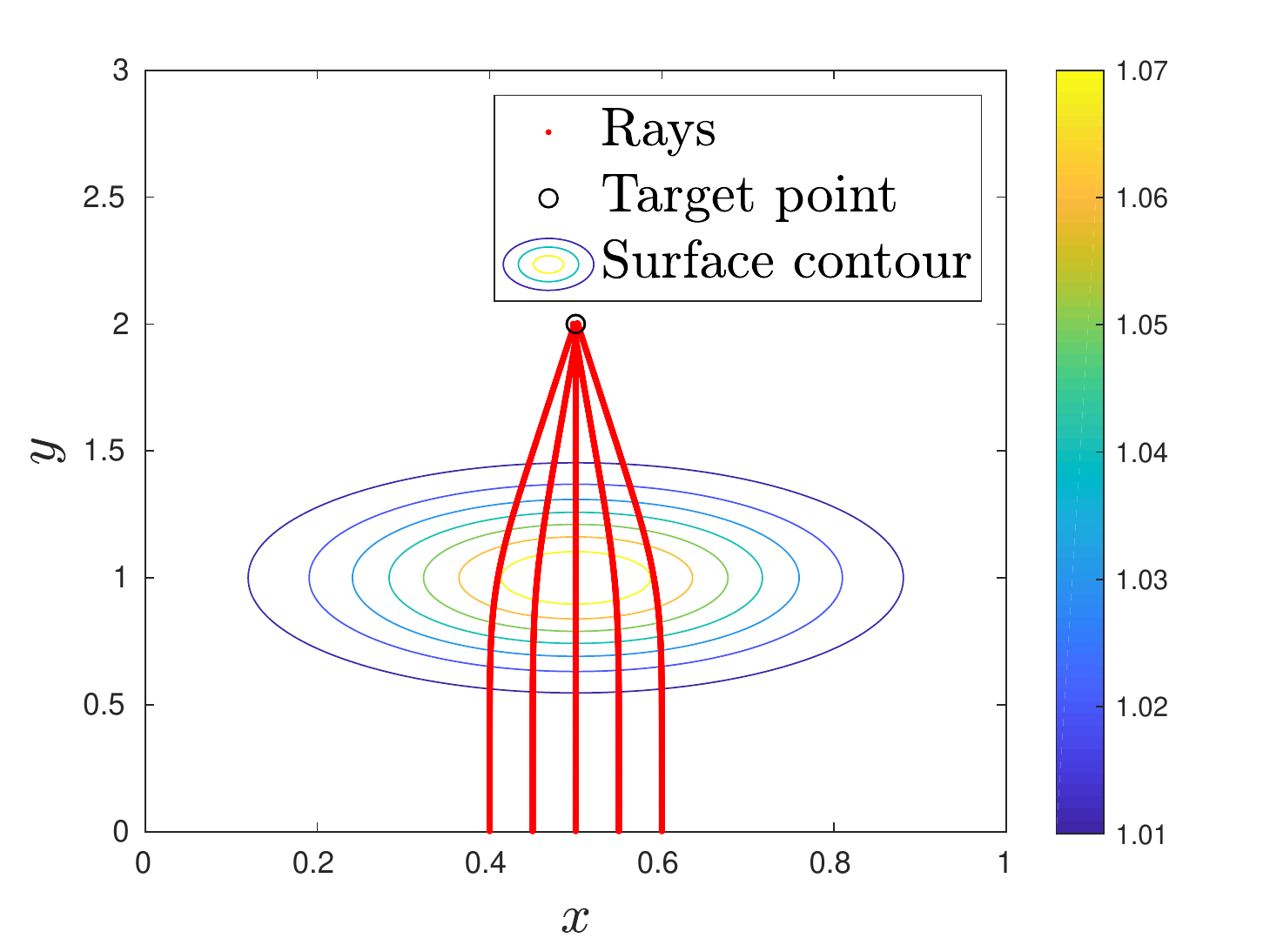}
\includegraphics[height=5cm,keepaspectratio]{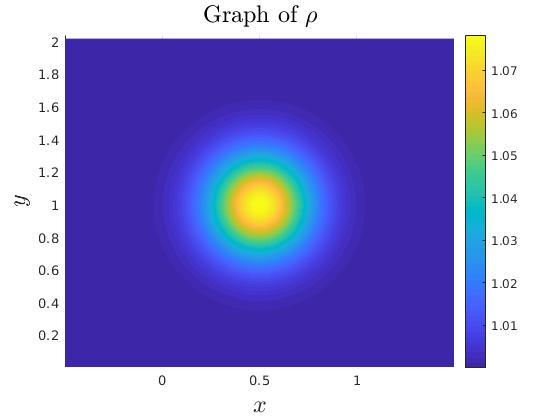}
\end{center}
\caption{ (Left). Crossing rays. (Right)  Graph of optimized $\rho_{\pi^*}$.}
\label{fig5}
\end{figure}

{\it Step 1.} The first step is to search for a target function $\rho_T = n^{2}_{T}$ parameterized with $\pi=(A,w_x,w_y)$ and $\sigma$, such that
\begin{eqnarray*}
\rho_{T}({\boldsymbol x})
&=& 1+A\exp\big(-w_x(x-0.5)^2-w_y(y-1)^2\big) \, .
\end{eqnarray*}
The target point where the trajectories are crossing is ${\boldsymbol x}_T=(0.5,2)$ and the $4$-dimensional search space is given by $\big(\pi,N_{\sigma}\big) \in  [0.05,0.15]\times[2,20]\times[2,20]\times[1.9,2.1]$.  When one of the stopping criteria is reached, the genetic algorithm provides the following champion $(\pi^{*},\sigma^{*})=(0.0779,14.1926,9.9920,1.9959)$. We report in Fig. \ref{fig5} (Left), the electron-rays crossing at the target and the corresponding function $\rho_{\pi^*}$.
 
\begin{figure}[hbt!]
\begin{center}
\includegraphics[height=3.5cm,keepaspectratio]{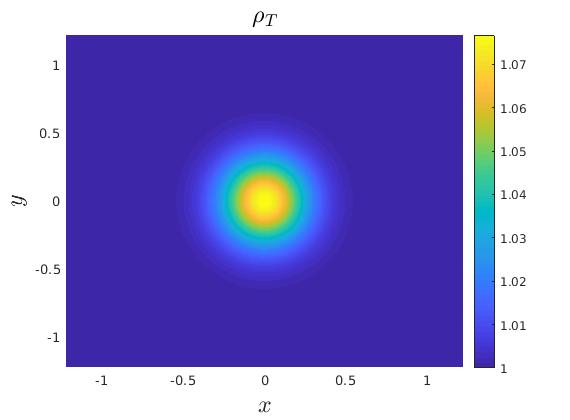}
\includegraphics[height=3.5cm,keepaspectratio]{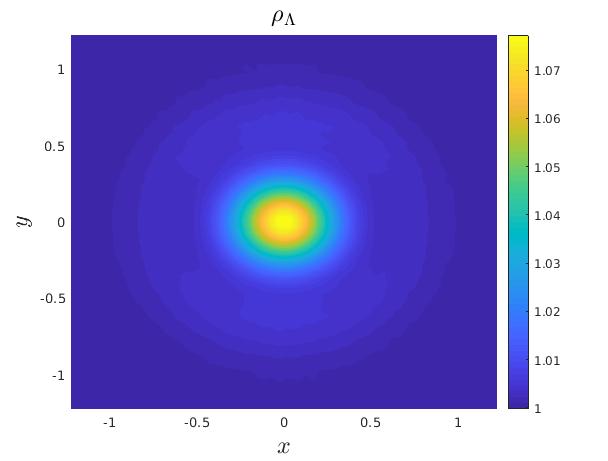}
\includegraphics[height=3.5cm,keepaspectratio]{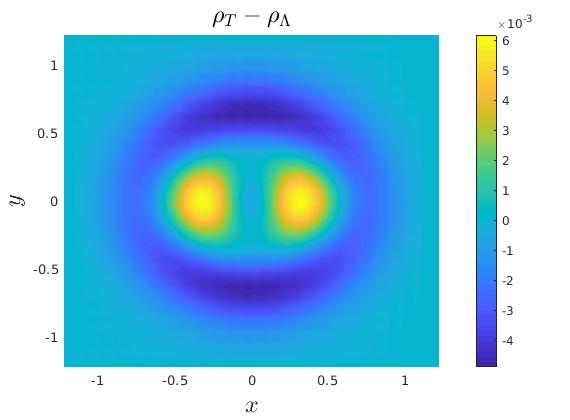}
\end{center}
\caption{(Left) Graph of the target $\rho_T$. (Right) Graph of the optimized $\rho_{\Lambda^{*}}$. (Right) Error: $\rho_{T}-\rho_{\Lambda}$ }
\label{fig8}
\end{figure}
\begin{figure}[hbt!]
\begin{center}
\includegraphics[height=8cm,keepaspectratio]{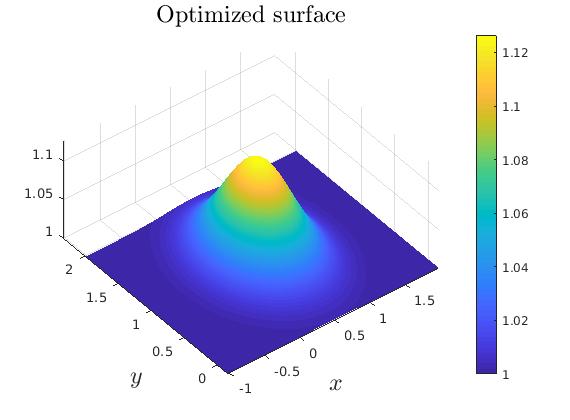}
\end{center}
\caption{{\bf Experiment 5.} (Left) Graph of optimized surface.}
\label{fig9}
\end{figure}

{\it Step 2.} Next, we next want to determine the surface $\mathcal{S}$ that will reproduce $\rho_{\pi^*}$ obtained in step 1. We choose an out-of-plane Gaussian deformation where the surface is $\mathcal{S}_{\Lambda^*}=\{({\boldsymbol x},Z_{\Lambda}({\boldsymbol x}))\}$, with
\begin{eqnarray}
Z_{\Lambda}(\boldsymbol{x}) = A\exp\big(-w_x(x-0.5)^2-w_y(y-1)^2\big).
\end{eqnarray}
Again, the optimization parameters are $\Lambda = (A,w_x,w_y)$ and the search space $[0.05,0.2]\times[2,10]\times[2,10]$. The optimization algorithm provides the champion: $\Lambda^*=(A^*,w^*_x,w^*_y)=(0.1293,3.3860,2.9065)$. The index of refraction is displayed in Fig. \ref{fig8} and the surface $Z_{\Lambda^{*}}$ is reported in Fig. \ref{fig9}. Again, we can see that our approach allows us to reproduce the desired index of refraction and to control the behavior of trajectories.

\begin{figure}[hbt!]
\begin{center}
\includegraphics[height=12cm,keepaspectratio]{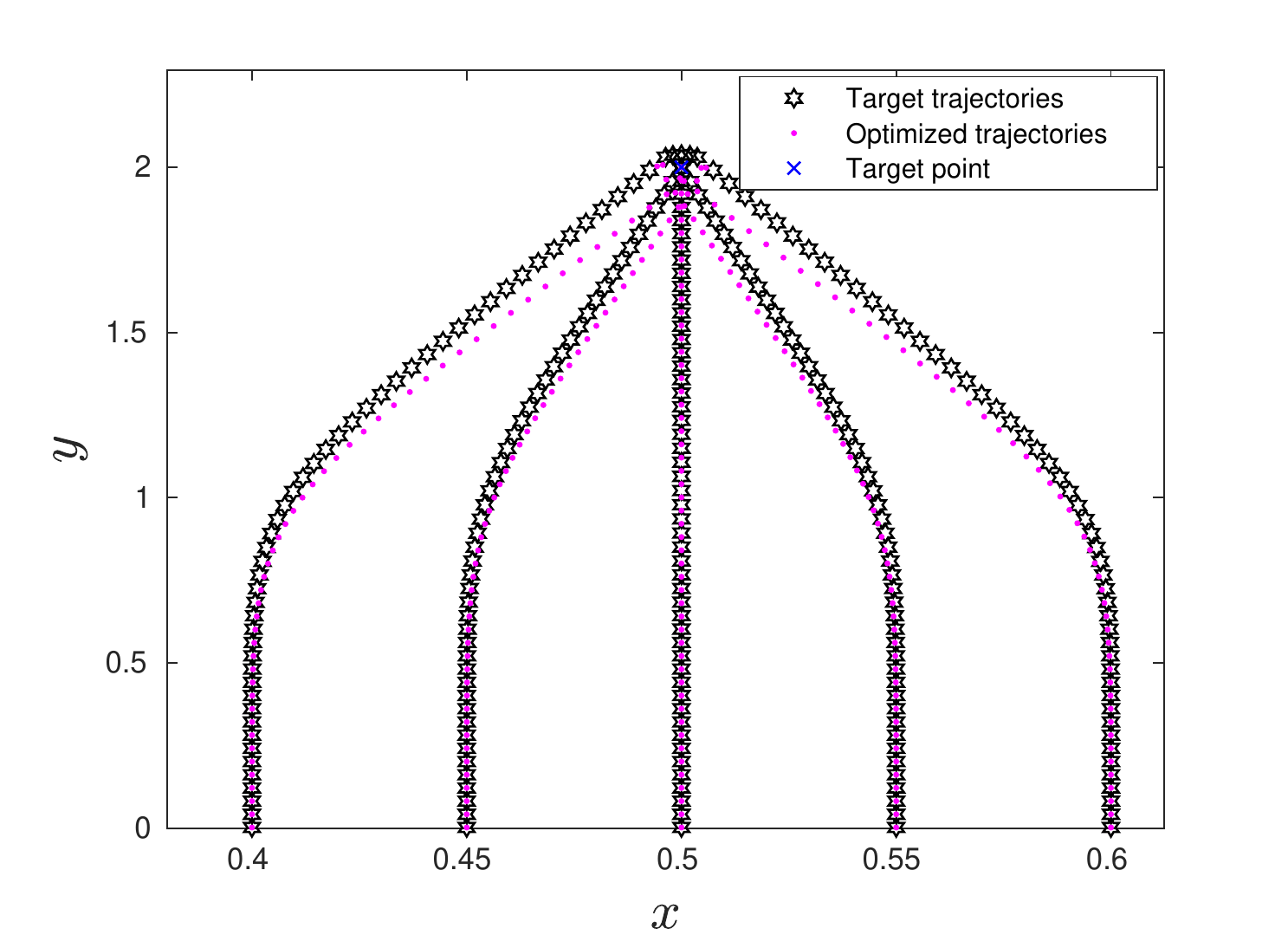}
\end{center}
\caption{Target and optimized trajectories.}
\label{fig10}
\end{figure}

Finally, in Fig. \ref{fig10}, we compare the electron-rays obtained in Step 1 (with $\rho_T$) to the ones on the optimized surface obtained in Step 2 (with $\rho_{\Lambda^*}$).  We conclude that the surface which was parameterized in Step 2 indeed possesses the searched refractive properties.

\subsection{Electron control}
In the following, we propose a simple test in which an electronic ray is guided from a known initial point to a final target point using 2 Gaussian surfaces.

\begin{figure}[hbt!]
\begin{center}
\includegraphics[height=6cm,keepaspectratio]{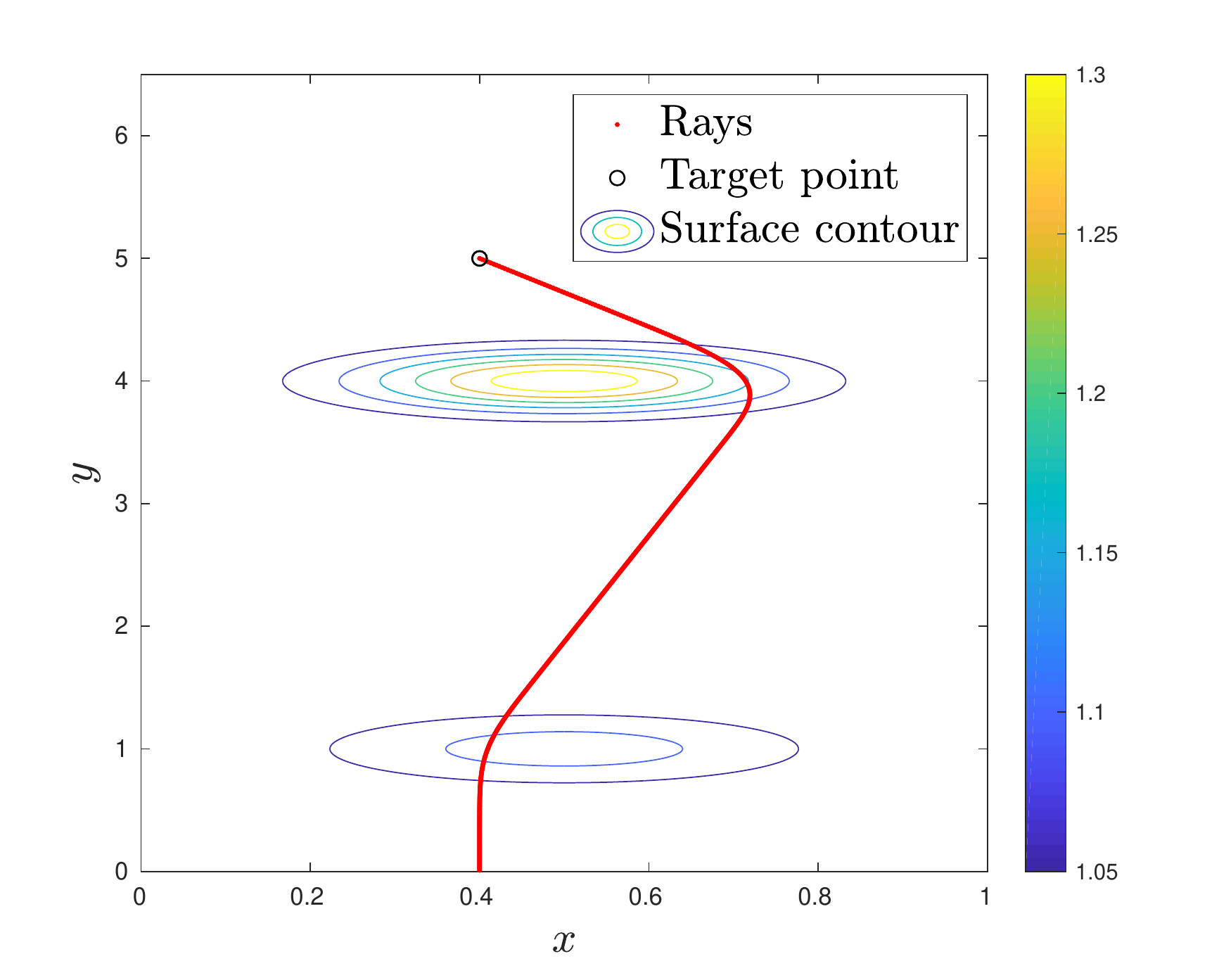}
\includegraphics[height=6cm,keepaspectratio]{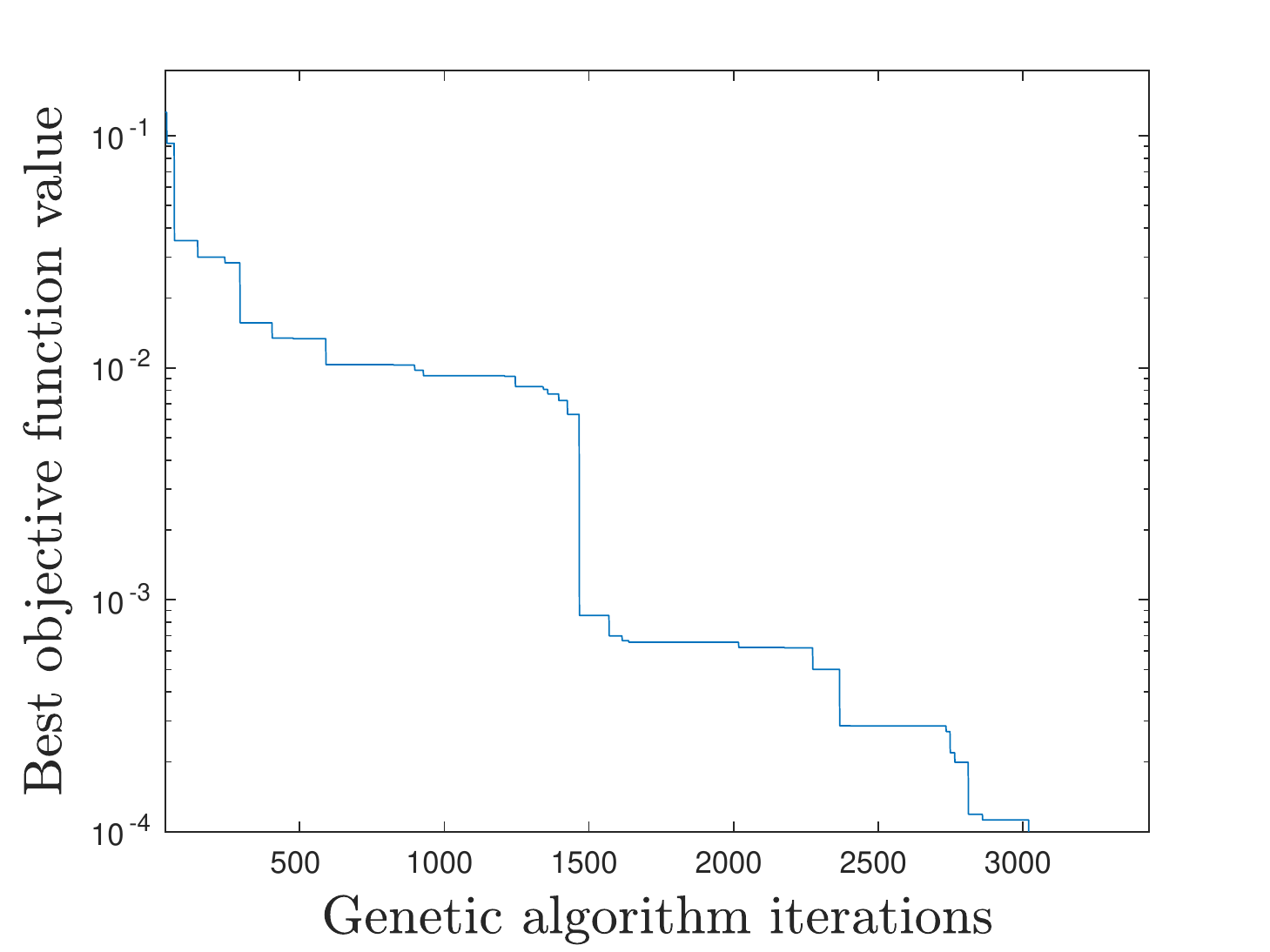}
\end{center}
\caption{ (Left). Crossing rays. (Right) Best discrete objective function as a function of total number of iterations of the genetic algorithm.}
\label{fig11}
\end{figure}

\noindent{\it Step 1.} A ray initially located in ${\boldsymbol x}_0=(0.4,0)$ is guided to ${\boldsymbol x}_T=(0.4,5)$ on a surface constituted by two Gaussian surfaces centered in $(0.5,1)$ and $(0.5,4)$. The widths $(w_1,w_2)$ and the amplititudes $(A_1,A_2)$ of the Gaussian surfaces are numerically optimized as well as the stepping,  using the same approach as before. The functional form of the desired index of refraction is chosen as
\begin{eqnarray*}
\rho_{\pi}({\boldsymbol x})  = 1+A_1\exp\big(-w_1(x-0.5)^2-w_1(y-1)^2\big) + A_2\exp\big(-w_2(x-0.5)^2-w_2(y-4)^2\big)\, 
\end{eqnarray*}
After optimization, the champion is given by\\
 $(w^*_1,A^*_1,w^*_2,A^*_2,\sigma^*)=(12.1517,0.1269,17.3673,0.3416,4.9467)$, see Fig. \ref{fig11}.

\begin{figure}[hbt!]
\begin{center}
\includegraphics[height=6cm,keepaspectratio]{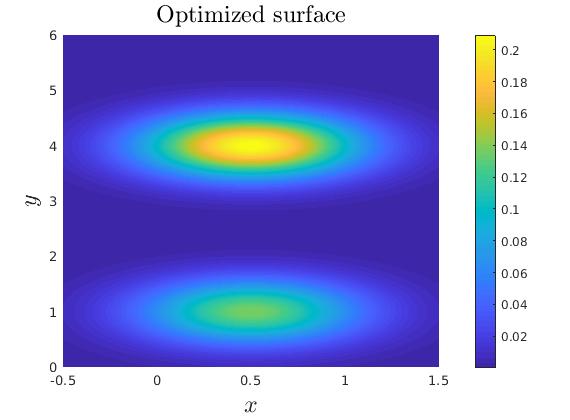}
\includegraphics[height=6cm,keepaspectratio]{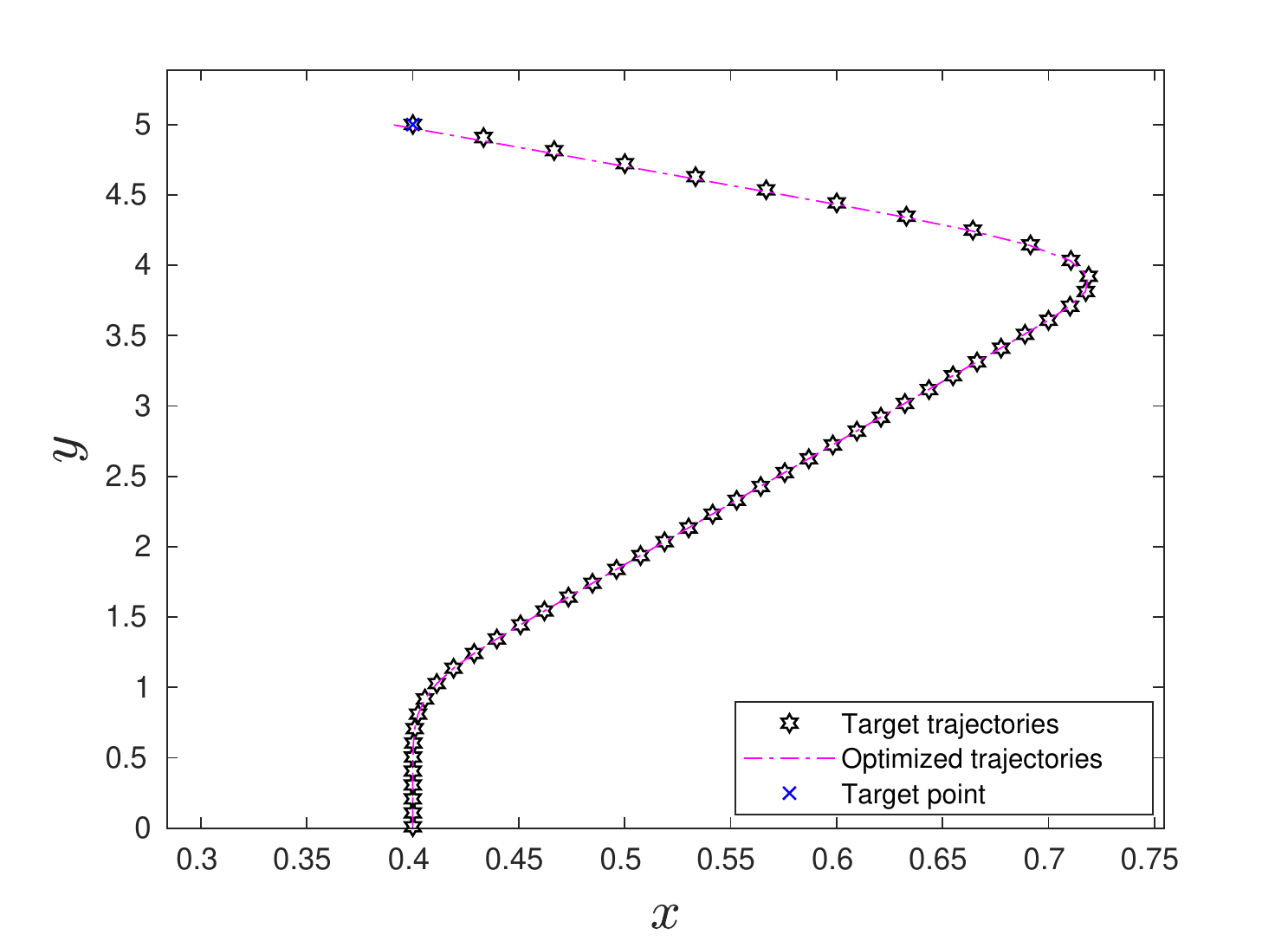}
\end{center}
\caption{ (Left) Graph of optimized surface. (Right.) Target and optimized trajectories.}
\label{fig12}
\end{figure}

{\it Step 2.} The next step is then to parameterize the surface. The out-of-plane deformation is chosen as
\begin{eqnarray*}
	Z_{\Lambda}({\boldsymbol x})  = A_1\exp\big(-w_1(x-0.5)^2-w_1(y-1)^2\big) + A_2\exp\big(-w_2(x-0.5)^2-w_2(y-4)^2\big)\, 
\end{eqnarray*}
The genetic algorithm provides the following optimized values: \\
$(w^*_1,A_1^*,w^*_2,A^*_2)=(4.2651,0.1634,3.6072,0.2375)$. 
The graph of the optimized surface is reported in Fig. \ref{fig12} (Left).

Finally, we compare the guided trajectory from Step 1 (from $\rho_\pi$) on the optimized surface obtained in Step 2. This surface has refractive index characterized by $\rho_{\Lambda^*}$.  We display in Fig. \ref{fig12} (Right), the target trajectory using $\rho_{\pi^*}$ computed in Step 1, and the optimized one using $\rho_{\Lambda^*}$ as computed in Step 2. This shows that the method is working well, up to a certain accuracy (related to the finite volume method accuracy, convergence of the optimization algorithm, etc).

\section{Conclusion}\label{sec:conclusion}
In this paper, we have developed and analyzed a general strategy to design graphene surfaces to control electron trajectories. The cornerstone of our approach is the use of isothermal coordinates, for which the metric tensor is diagonal and related to a graded refractive index in the semiclassical limit. Working in isothermal coordinates however requires the solution of the Beltrami equation, which is efficiently and accurately performed in this paper, thanks to a least-square finite volume method. By combining this numerical scheme to optimization algorithm, we have demonstrated that it is possible to inverse engineer a strained graphene surface with some desired scattering properties.
Some numerical experiments have shown the accuracy of our approach to parameterize graphene surface with refractive optics-like properties. Simple surface parametrizations have been considered to construct aberation-free lenses and to control electrons. Obviously, more intricate configurations are possible, which may allow to design refractive optical-like elements. This may be important to the development of new nanoelectronic devices.

Our optimization technique could be also be extended to more accurate models. For example, going beyond the semi-classical approximation is possible, in principle. To reach this goal, one possibility is to consider Bohm-like trajectories \cite{bohm1,bohm2} and optimize these trajectories on some objective. However, this requires a full solution of the Dirac equation (possibly with numerical methods presented in \cite{jcp2020,cpc2017}), which is computationally much more expensive than solving the classical equation of motion inside the graded index of refraction.

\bibliographystyle{unsrt}
\bibliography{refs}

\begin{thebibliography}{10}

\bibitem{Feng2016}
C.~Si, Z.~Sun, and F.~Liu.
\newblock Strain engineering of graphene: a review.
\newblock {\em Nanoscale}, 8:3207--3217, 2016.

\bibitem{GUINEA20121437}
F.~Guinea.
\newblock Strain engineering in graphene.
\newblock {\em Solid State Communications}, 152(15):1437 -- 1441, 2012.
\newblock Exploring Graphene, Recent Research Advances.

\bibitem{PhysRevLett.103.046801}
V.~M. Pereira and A.~H. Castro~N.
\newblock Strain engineering of graphene's electronic structure.
\newblock {\em Phys. Rev. Lett.}, 103:046801, Jul 2009.

\bibitem{Naumis_2017}
G.~G. Naumis, S.~Barraza-Lopez, M.~Oliva-Leyva, and H.~Terrones.
\newblock Electronic and optical properties of strained graphene and other
  strained 2d materials: a review.
\newblock {\em Reports on Progress in Physics}, 80(9):096501, aug 2017.

\bibitem{AMORIM20161}
B.~Amorim, A.~Cortijo, F.~[de Juan], A.G. Grushin, F.~Guinea,
  A.~Gutiérrez-Rubio, H.~Ochoa, V.~Parente, R.~Roldán, P.~San-Jose,
  J.~Schiefele, M.~Sturla, and M.A.H. Vozmediano.
\newblock Novel effects of strains in graphene and other two dimensional
  materials.
\newblock {\em Physics Reports}, 617:1 -- 54, 2016.

\bibitem{PhysRevB.81.081407}
S.-M. Choi, S.-H. Jhi, and Y.-W. Son.
\newblock Effects of strain on electronic properties of graphene.
\newblock {\em Phys. Rev. B}, 81:081407, Feb 2010.

\bibitem{boggild2017two}
Peter B{\o}ggild, Jos{\'e}~M Caridad, Christoph Stampfer, Gaetano Calogero,
  Nick~R{\"u}bner Papior, and Mads Brandbyge.
\newblock A two-dimensional dirac fermion microscope.
\newblock {\em Nature communications}, 8(1):1--12, 2017.

\bibitem{PhysRevB.76.165409}
F.~de~Juan, A.~Cortijo, and M.~A.~H. Vozmediano.
\newblock Charge inhomogeneities due to smooth ripples in graphene sheets.
\newblock {\em Phys. Rev. B}, 76:165409, Oct 2007.

\bibitem{PhysRevLett.108.227205}
F.~de~Juan, M.~Sturla, and M.~A.~H. Vozmediano.
\newblock Space dependent fermi velocity in strained graphene.
\newblock {\em Phys. Rev. Lett.}, 108:227205, May 2012.

\bibitem{PhysRevB.87.165131}
F.~de~Juan, J.~L. Ma\~nes, and M.~A.~H. Vozmediano.
\newblock Gauge fields from strain in graphene.
\newblock {\em Phys. Rev. B}, 87:165131, Apr 2013.

\bibitem{OLIVALEYVA20152645}
M.~Oliva-Leyva and Gerardo~G. Naumis.
\newblock Generalizing the fermi velocity of strained graphene from uniform to
  nonuniform strain.
\newblock {\em Physics Letters A}, 379(40):2645--2651, 2015.

\bibitem{VOLOVIK2014352}
G.E. Volovik and M.A. Zubkov.
\newblock Emergent horava gravity in graphene.
\newblock {\em Annals of Physics}, 340(1):352 -- 368, 2014.

\bibitem{PhysRevB.82.073405}
A.~Mesaros, D.~Sadri, and J.~Zaanen.
\newblock Parallel transport of electrons in graphene parallels gravity.
\newblock {\em Phys. Rev. B}, 82:073405, Aug 2010.

\bibitem{VOZMEDIANO2010109}
M.A.H. Vozmediano, M.I. Katsnelson, and F.~Guinea.
\newblock Gauge fields in graphene.
\newblock {\em Physics Reports}, 496(4):109 -- 148, 2010.

\bibitem{Gallerati2019}
A.~Gallerati.
\newblock Graphene properties from curved space {D}irac equation.
\newblock {\em European Physical Journal Plus}, 134(5), 2019.

\bibitem{PhysRevB.81.035408}
F.~Guinea, A.~K. Geim, M.~I. Katsnelson, and K.~S. Novoselov.
\newblock Generating quantizing pseudomagnetic fields by bending graphene
  ribbons.
\newblock {\em Phys. Rev. B}, 81:035408, Jan 2010.

\bibitem{Debus_2018}
J-D Debus, M~Mendoza, and H~J Herrmann.
\newblock Shifted landau levels in curved graphene sheets.
\newblock {\em Journal of Physics: Condensed Matter}, 30(41):415503, sep 2018.

\bibitem{PhysRevB.95.125432}
Pavel Castro-Villarreal and R.~Ruiz-S\'anchez.
\newblock Pseudomagnetic field in curved graphene.
\newblock {\em Phys. Rev. B}, 95:125432, Mar 2017.

\bibitem{PhysRevB.84.081401}
Kyung-Joong Kim, Ya.~M. Blanter, and Kang-Hun Ahn.
\newblock Interplay between real and pseudomagnetic field in graphene with
  strain.
\newblock {\em Phys. Rev. B}, 84:081401, Aug 2011.

\bibitem{RAMEZANIMASIR201376}
M.~{Ramezani Masir}, D.~Moldovan, and F.M. Peeters.
\newblock Pseudo magnetic field in strained graphene: Revisited.
\newblock {\em Solid State Communications}, 175-176:76 -- 82, 2013.
\newblock Special Issue: Graphene V: Recent Advances in Studies of Graphene and
  Graphene analogues.

\bibitem{Levy544}
N.~Levy, S.~A. Burke, K.~L. Meaker, M.~Panlasigui, A.~Zettl, F.~Guinea,
  A.~H.~Castro Neto, and M.~F. Crommie.
\newblock Strain-induced pseudo{\textendash}magnetic fields greater than 300
  {T}esla in graphene nanobubbles.
\newblock {\em Science}, 329(5991):544--547, 2010.

\bibitem{CLY}
L.~Chai, E.~Lorin, and X.~Yang.
\newblock Frozen gaussian approximation for the dirac equation in curved space
  with application to strained graphene.
\newblock {\em Submitted}, 2021.

\bibitem{Chaves_2014}
A.~J. Chaves, T.~Frederico, O.~Oliveira, W.~de~Paula, and M.~C. Santos.
\newblock Optical conductivity of curved graphene.
\newblock {\em Journal of Physics: Condensed Matter}, 26(18):185301, apr 2014.

\bibitem{Contreras_Astorga_2020}
A.~Contreras-Astorga, V.~Jakubsk{\'{y}}, and A.~Raya.
\newblock On the propagation of {D}irac fermions in graphene with
  strain-induced inhomogeneous {F}ermi velocity.
\newblock {\em Journal of Physics: Condensed Matter}, 32(29):295301, apr 2020.

\bibitem{PhysRevB.98.155419}
Kyriakos Flouris, Miller Mendoza~Jimenez, Jens-Daniel Debus, and Hans~J.
  Herrmann.
\newblock Confining massless dirac particles in two-dimensional curved space.
\newblock {\em Phys. Rev. B}, 98:155419, Oct 2018.

\bibitem{Stegmann_2016}
Thomas Stegmann and Nikodem Szpak.
\newblock Current flow paths in deformed graphene: from quantum transport to
  classical trajectories in curved space.
\newblock {\em New Journal of Physics}, 18(5):053016, may 2016.

\bibitem{PhysRevE.103.013312}
F.~Fillion-Gourdeau, E.~Lorin, and S.~MacLean.
\newblock Numerical quasiconformal transformations for electron dynamics on
  strained graphene surfaces.
\newblock {\em Phys. Rev. E}, 103:013312, Jan 2021.

\bibitem{semiclassical2021}
F.~Fillion-Gourdeau, E.~Lorin, and S.~Maclean.
\newblock Two-dimensional dirac matter in the semiclassical regime.
\newblock 2021.
\newblock arXiv:2111.11496.

\bibitem{Cortijo_2007}
A.~Cortijo and M.~A.~H. Vozmediano.
\newblock Electronic properties of curved graphene sheets.
\newblock {\em Europhysics Letters ({EPL})}, 77(4):47002, feb 2007.

\bibitem{Vozmediano_2008}
M.~A.~H. Vozmediano, F.~de~Juan, and A.~Cortijo.
\newblock Gauge fields and curvature in graphene.
\newblock {\em Journal of Physics: Conference Series}, 129:012001, oct 2008.

\bibitem{PhysRevB.88.085430}
M.~Oliva-Leyva and Gerardo~G. Naumis.
\newblock Understanding electron behavior in strained graphene as a reciprocal
  space distortion.
\newblock {\em Phys. Rev. B}, 88:085430, Aug 2013.

\bibitem{PhysRevB.88.155405}
Juan~L. Ma\~nes, Fernando de~Juan, Mauricio Sturla, and Mar\'{\i}a A.~H.
  Vozmediano.
\newblock Generalized effective hamiltonian for graphene under nonuniform
  strain.
\newblock {\em Phys. Rev. B}, 88:155405, Oct 2013.

\bibitem{pollock2010dirac}
M.D. Pollock.
\newblock On the dirac equation in curved space-time.
\newblock {\em Acta Physica Polonica B}, 41(8), 2010.

\bibitem{PhysRevB.92.245110}
Enrique Arias, Alexis~R. Hern\'andez, and Caio Lewenkopf.
\newblock Gauge fields in graphene with nonuniform elastic deformations: A
  quantum field theory approach.
\newblock {\em Phys. Rev. B}, 92:245110, Dec 2015.

\bibitem{ahlfors2006lectures}
L.~V. Ahlfors.
\newblock {\em Lectures on quasiconformal mappings}, volume~38.
\newblock American Mathematical Soc., 2006.

\bibitem{maslov2001semi}
Victor~P Maslov and Mikhail~Vasilevich Fedoriuk.
\newblock {\em Semi-classical approximation in quantum mechanics}, volume~7.
\newblock Springer Science \& Business Media, 2001.

\bibitem{Newton}
J.~Evans and M.~Rosenquist.
\newblock “f—m a” optics.
\newblock {\em American Journal of Physics}, 54(10):876--883, 1986.

\bibitem{BARTELMANN2001291}
M.~Bartelmann and P.~Schneider.
\newblock Weak gravitational lensing.
\newblock {\em Physics Reports}, 340(4):291 -- 472, 2001.

\bibitem{PhysRevB.82.205430}
A.~Chaves, L.~Covaci, Kh.~Yu. Rakhimov, G.~A. Farias, and F.~M. Peeters.
\newblock Wave-packet dynamics and valley filter in strained graphene.
\newblock {\em Phys. Rev. B}, 82:205430, Nov 2010.

\bibitem{PhysRevB.78.235321}
G.~M. Maksimova, V.~Ya. Demikhovskii, and E.~V. Frolova.
\newblock Wave packet dynamics in a monolayer graphene.
\newblock {\em Phys. Rev. B}, 78:235321, Dec 2008.

\bibitem{doi:10.1021/nn800459e}
Z.~Hua Ni, T.~Yu, Y.H. Lu, Ying~Y. Wang, Y.~P. Feng, and Z.~X. Shen.
\newblock Uniaxial strain on graphene: Raman spectroscopy study and band-gap
  opening.
\newblock {\em ACS Nano}, 2(11):2301--2305, 2008.
\newblock PMID: 19206396.

\bibitem{RevModPhys.81.109}
A.~H. Castro~Neto, F.~Guinea, N.~M.~R. Peres, K.~S. Novoselov, and A.~K. Geim.
\newblock The electronic properties of graphene.
\newblock {\em Rev. Mod. Phys.}, 81:109--162, Jan 2009.

\bibitem{LEONHARDT200969}
Ulf Leonhardt and Thomas~G. Philbin.
\newblock Chapter 2 transformation optics and the geometry of light.
\newblock volume~53 of {\em Progress in Optics}, pages 69--152. Elsevier, 2009.

\bibitem{ga1}
D.~E. Goldberg.
\newblock {\em The design of innovation}, volume~7 of {\em Genetic Algorithms
  and Evolutionary Computation}.
\newblock Kluwer Academic Publishers, Boston, MA, 2002.
\newblock Lessons from and for competent genetic algorithms.

\bibitem{ga2}
A.~R. Conn, N.~I.~M. Gould, and P.~L. Toint.
\newblock A globally convergent augmented {L}agrangian algorithm for
  optimization with general constraints and simple bounds.
\newblock {\em SIAM J. Numer. Anal.}, 28(2):545--572, 1991.

\bibitem{yang2010nature}
Xin-She Yang.
\newblock {\em Nature-inspired metaheuristic algorithms}.
\newblock Luniver press, Frome, United Kingdom, 2010.

\bibitem{MR1925043}
R.~J. LeVeque.
\newblock {\em Finite volume methods for hyperbolic problems}.
\newblock Cambridge Texts in Applied Mathematics. Cambridge University Press,
  Cambridge, 2002.

\bibitem{Lee385}
Changgu Lee, Xiaoding Wei, Jeffrey~W. Kysar, and James Hone.
\newblock Measurement of the elastic properties and intrinsic strength of
  monolayer graphene.
\newblock {\em Science}, 321(5887):385--388, 2008.

\bibitem{bohm1}
P.~Bracken.
\newblock Metric geometry and the determination of the bohmian quantum
  potential.
\newblock {\em Journal of Physics Communications}, 3(6), 2019.

\bibitem{bohm2}
C.~Colijn and E.R. Vrscay.
\newblock Spin-dependent bohm trajectories for pauli and dirac eigenstates of
  hydrogen.
\newblock {\em Foundations of Physics Letters}, 16(4):303--323, 2003.

\bibitem{jcp2020}
X.~Antoine, F.~Fillion-Gourdeau, E.~Lorin, and S.~MacLean.
\newblock Pseudospectral computational methods for the time-dependent {D}irac
  equation in static curved spaces.
\newblock {\em J. of Comput. Phys.}, 411:109412, 2020.

\bibitem{cpc2017}
X.~Antoine and E.~Lorin.
\newblock Computational performance of simple and efficient sequential and
  parallel {D}irac equation solvers.
\newblock {\em Comput. Phys. Commun.}, 220:150--172, 2017.

\end{thebibliography}

\end{document}